%% file: main.tex
\documentclass[sigconf, nonacm]{acmart}

\AtBeginDocument{%
  }

\input{packages}

\newcommand{\para}[1]{\noindent\textbf{{#1}}}
\def\ourmethod{\textsc{Misrouter}}

\begin{document}

\title{\ourmethod{}: Exploiting Routing Mechanisms for Input-Only Attacks on Mixture-of-Experts LLMs}

\author{Zekun Fei}
\authornote{The first two authors contributed equally to this work.}
\affiliation{
  \institution{Nankai University}
  \city{Tianjin}
  \country{China}
}
\email{feizekun@mail.nankai.edu.cn}

\author{Zihao Wang}
\authornotemark[1]
\authornote{Correspondence to \href{mailto:zihao.wang@ntu.edu.sg}{zihao.wang@ntu.edu.sg}.}
\affiliation{
  \institution{Nanyang Technological University}
  \country{Singapore}
}
\email{zihao.wang@ntu.edu.sg}

\author{Weijie Liu}
\affiliation{
  \institution{Nankai University}
  \city{Tianjin}
  \country{China}
}
\email{weijieliu@nankai.edu.cn}

\author{Ruiqi He}
\affiliation{
  \institution{Nankai University}
  \city{Tianjin}
  \country{China}
}
\email{heruiqi@mail.nankai.edu.cn}

\author{Jianing Geng}
\affiliation{
  \institution{Nankai University}
  \city{Tianjin}
  \country{China}
}
\email{gengjianing@mail.nankai.edu.cn}

\author{Zheli Liu}
\affiliation{
  \institution{Nankai University}
  \city{Tianjin}
  \country{China}
}
\email{liuzheli@nankai.edu.cn}

\author{XiaoFeng Wang}
\affiliation{
  \institution{Nanyang Technological University}
  \country{Singapore}
}
\email{xiaofeng.wang@ntu.edu.sg}

\date{}

\input{0_abstract.tex}

\maketitle

\input{1_introduction}

\input{2_background}
\input{3_method}

\input{4_evaluation}

\input{5_discussion}

\input{6_conclusion}

\input{main.bbl}

\input{99_appendix}

\end{document}

%% file: packages.tex
\usepackage{tikz}
\usepackage{amsthm}
\usepackage{autobreak}
\allowdisplaybreaks
\usepackage{adjustbox}
\usepackage{booktabs}
\usepackage{makecell}
\usepackage{siunitx}
\usepackage{multirow}
\usepackage{balance}
\usepackage{longtable}
\usepackage{algorithm}
\usepackage[algo2e]{algorithm2e}
\usepackage{algorithmic}
\usepackage{bm}
\usepackage{graphicx}
\usepackage{placeins}
\usepackage{amsmath}

\usepackage{booktabs}
\usepackage{tabularx}
\usepackage{listings}
\usepackage{fix-cm}
\usepackage{hyperref}
\usepackage{wasysym}

\usepackage{amssymb}
\usepackage{array}
\usepackage{tabularx}
\usepackage{caption}
\usepackage{xcolor}
\usepackage{bm}
\usepackage{amsthm}

\usepackage{enumitem}
\usepackage{threeparttable}
\usepackage{float}

\usepackage[framemethod=tikz]{mdframed}
\usepackage{colortbl}
\usepackage{xcolor,soul}
\definecolor{lightyellow}{RGB}{255,255,153}
\definecolor{lightblue}{RGB}{153,255,255}

\mdfdefinestyle{highlighted}{%
    linecolor=white,          
    linewidth=0pt,          
    backgroundcolor=lightyellow, 
    innerleftmargin=2pt,   
    innerrightmargin=2pt,  
    innertopmargin=4pt,    
    innerbottommargin=4pt, 
}

\newmdenv[style=highlighted]{highlightedparagraph}
\newcommand{\pgftextcircled}[1]{
    \setbox0=\hbox{#1}%
    \dimen0\wd0%
    \divide\dimen0 by 2%
    \begin{tikzpicture}[baseline=(a.base)]%
        \useasboundingbox (-\the\dimen0,0pt) rectangle (\the\dimen0,1pt);
        \node[circle,draw,outer sep=0pt,inner sep=0.1ex] (a) {#1};
    \end{tikzpicture}
}

\sethlcolor{lightyellow}

\colorlet{highlightblue}{cyan!30}

\theoremstyle{definition}
\theoremstyle{remark}

\newtheoremstyle{bfnote}%
  {}{}
  {\itshape}{}
  {\bfseries}{.}
  { }{\thmname{#1}\thmnumber{ #2}\thmnote{ (#3)}}
\theoremstyle{bfnote}

\hypersetup{
  colorlinks,
  linkcolor={blue!70!green},
  citecolor={green!70!blue},
  urlcolor={orange!70!red}
}

%% file: 0_abstract.tex
\begin{abstract}
Mixture-of-Experts (MoE) architectures have emerged as a leading paradigm for scaling large language models through sparse, routing-based computation. However, this design also introduces a new attack surface: the routing mechanism that determines which experts process each input. Prior work shows that manipulating routing can bypass safety alignment, but existing attacks require model modification and thus apply only to locally deployed models. By contrast, real-world LLM services are remotely hosted and accessible only through input queries. This raises a fundamental question: can MoE routing be exploited through input-only attacks to induce stronger unsafe behaviors in real-world services?

Our key insight is to optimize attacks in a white-box setting on open-source surrogate MoE models and transfer the resulting adversarial inputs to public API services within the same model family. This setting presents three main challenges: routing can be influenced only indirectly through input perturbations, routing control and output generation are tightly coupled, and even a successful safety bypass may still produce low-quality responses.

To address these challenges, we propose \ourmethod{}, an input-only attack framework that jointly targets routing behavior and expert functionality. Rather than simply avoiding strongly aligned experts, \ourmethod{} identifies weakly aligned experts that are willing to produce target harmful content by analyzing expert activations under harmful queries paired with unsafe continuations. It then optimizes adversarial inputs to steer routing toward these experts and away from strongly aligned ones. To maintain output quality, it further biases routing toward highly capable general-purpose experts identified from benign question-answering tasks. Finally, because routing and output objectives can conflict, \ourmethod{} uses a two-phase optimization strategy that first steers routing and then optimizes harmful outputs while preserving routing stability. 
\end{abstract}

%% file: 1_introduction.tex
\section{Introduction}
Mixture-of-Experts (MoE)~\cite{jacobs1991adaptive, shazeer2017outrageously, du2022glam, cai2025survey} has emerged as a leading paradigm for scaling large language models through sparse, routing-based computation. Unlike dense models, MoE architectures rely on dynamic expert selection, where only a subset of experts is activated for each input. While this design improves scaling efficiency, it also introduces a new attack surface: the routing mechanism that determines which experts process each input.

Recent work has shown that manipulating routing can bypass safety alignment in MoE models~\cite{jiang2026sparse, fayyaz2025steering, wu2025gatebreaker, lai2025safex}. However, existing routing-based attacks typically require direct model modification, such as altering routing logits or expert computations. As a result, they apply only to locally deployed models. By contrast, real-world LLM services are remotely hosted and accessible only through input queries, where attackers cannot inspect or modify the model.

In this work, we study \emph{input-only attacks} against MoE LLMs, where the attacker manipulates only the input without modifying the model. This setting naturally applies to remotely hosted services and extends routing-based attacks beyond locally deployable models. This leads to the following research question: \emph{Can an attacker exploit the routing mechanism of MoE-based LLM services to induce unsafe behaviors using only input manipulation?}

\para{Challenges.}
A straightforward way to answer this question is to optimize adversarial inputs on an open-source surrogate MoE model and transfer them to public API services within the same model family. However, we find that this direct approach fails in practice (see \autoref{sec:eval}), due to the following challenges.

\textit{First}, routing can only be influenced indirectly through input perturbations rather than explicitly controlled. As a result, the attacker can only steer the model toward a desired routing pattern approximately, instead of enforcing it precisely. This limitation becomes especially severe when the desired routing behavior differs substantially from the model's default routing pattern, making reliable input-only steering difficult.

\textit{Second}, routing control and output generation are inherently coupled, and optimizing one objective may interfere with the other. Input changes that strengthen generation of harmful content can alter routing decisions, while constraints imposed to preserve a target routing pattern may hinder the production of unsafe outputs. This coupling creates an unstable optimization problem and can lead to suboptimal solutions.

\textit{Third}, even when safety mechanisms are bypassed, the resulting outputs may still be of poor quality. Changing routing decisions can direct the input to experts that are not well suited for the task, leading to responses that are vague, incoherent, or uninformative. For example, steering the model away from strongly aligned experts may reduce refusal behavior, but does not ensure that the selected experts can generate a high-quality answer to the target query. Therefore, a successful attack must not only bypass safety alignment, but also preserve sufficient output quality.

\para{Our solution.}
To address these challenges, we propose \ourmethod{}, an input-only attack framework that jointly targets routing behavior and expert functionality.
To mitigate the limited controllability of routing, \ourmethod{} moves beyond simply avoiding strongly aligned experts. Prior routing-based attacks~\cite{wu2025gatebreaker} identify strongly aligned experts by probing the model with harmful queries and observing which experts are activated when the model produces refusal-style continuations (e.g., ``I can't help with that...''). A straightforward input-only extension would therefore try to steer inputs away from these experts. However, this objective is often too rigid under input-only constraints: it is difficult to reliably avoid a specific set of experts using input perturbations alone. More importantly, bypassing strongly aligned experts does not necessarily activate experts that are willing to produce the target harmful content (see \autoref{subsec:mask}).
Instead, \ourmethod{} explicitly identifies weakly aligned experts that are willing to produce target harmful content. Specifically, we feed the model harmful queries paired with unsafe continuations and analyze the resulting expert activations. We then optimize adversarial inputs to steer routing toward these weakly aligned experts and away from strongly aligned ones. This provides a more attainable routing objective under input-only constraints and substantially improves attack effectiveness.

To address the optimization difficulty caused by coupled objectives, \ourmethod{} decouples routing control and output generation through a two-phase strategy. In the first phase, it optimizes the input to reach a feasible target routing pattern, without interference from output-related objectives. Once such a routing pattern is established, the attack effectively focuses on a more stable set of activated experts. In the second phase, \ourmethod{} jointly optimizes for harmful outputs while preserving routing stability. The routing objective acts as a constraint that prevents the optimized prompt from drifting away from the desired routing behavior.

Finally, to maintain output quality, \ourmethod{} biases routing toward highly capable general-purpose experts. Our observation is that although MoE models distribute computation across different experts, some experts are consistently activated by diverse benign question-answering tasks and are associated with stronger general generation capability (see \autoref{subsec:mask}). We identify these highly capable general-purpose experts using benign queries and encourage adversarial inputs to route through them. In this way, the attack not only reduces refusal behavior, but also preserves a basic level of response quality, which is critical in input-only settings where the attacker lacks fine-grained control over expert activations.

Together, these designs enable effective and stable input-only attacks on MoE-based LLMs, extending routing-based attacks beyond locally deployable models to remotely hosted services.

\para{Empirical results.}
We conduct extensive experiments under three settings: a white-box setting where the surrogate and target models are identical, a gray-box setting where attacks transfer from an open-weight base model to its fine-tuned variants, and a black-box setting where attacks transfer from open-source surrogate MoE models to public API services. In the black-box setting, the target API may differ from the surrogate in model scale, number of experts, layer depth, routing design, safety tuning, system prompts, and deployment-time filtering, making transfer substantially more challenging.
Our main evaluation covers six popular MoE LLM API services from five major developers, including OpenAI, Microsoft, DeepSeek, Alibaba, and Mistral, using two benchmark datasets, StrongREJECT~\cite{souly2024strongreject} and AdvBench~\cite{zou2023universal}. 

In the black-box transfer setting, the results show that routing-aware optimization provides additional attack capability beyond conventional input-only jailbreak strategies. We evaluate both standalone \ourmethod{} and a compositional variant, \ourmethod{}+FFA, which applies \ourmethod{} on top of FFA-generated prompts. This combined attack improves the average ASR over representative input-only jailbreak baselines from 20.8\% to 39.7\%. Note that we compute ASR using an LLM-based judge~\cite{inan2023llama}, which considers not only whether the model avoids refusal, but also whether the response contains sufficiently informative harmful content.

Additional analyses across the three settings provide evidence for both the mechanism and transferability of \ourmethod{}. In white-box experiments, \ourmethod{} consistently reduces routing loss, confirming that the performance gains stem from effective routing manipulation. In gray-box experiments, \ourmethod{} transfers effectively from base MoE models to their fine-tuned variants, suggesting that routing-aware prompts can remain effective when the target model preserves the same architecture and initialization. In black-box experiments, we find that standard output-level optimization, such as GCG~\cite{zou2023universal}, transfers poorly and often yields near-zero ASR. In contrast, \ourmethod{} shows stronger transferability because it targets the routing mechanism rather than the full output distribution. Intuitively, routing is a comparatively simpler and lower-dimensional decision component, whereas output generation depends on the combined behavior of many activated experts and is therefore more model-specific.

We further perform ablation studies to analyze the impact of routing target construction and optimization strategy. The results show that both the contrastive routing signal and the two-phase optimization are critical for achieving stable and effective attacks.

Finally, we explore potential countermeasures. We consider two defense directions. 
The first aims to eliminate weak routing patterns by strengthening under-protected components, so that no vulnerable routing configurations remain. However, this requires uniformly improving the robustness of the system, which incurs substantial overhead and runs counter to the design principle of MoE. 
The second amplifies the effects of safety-aligned experts by increasing their routing probability, thereby biasing the model toward safer behaviors. In the extreme case, this reduces to directing all inputs to these experts. While more efficient, this approach degrades model utility, as it interferes with the specialization of experts and affects overall performance. 
Our evaluation shows that both strategies fail to effectively mitigate \ourmethod{}. While these defenses are expected to incur additional costs in efficiency or model utility, even with these costs, they remain insufficient to mitigate the attack. These findings suggest that \ourmethod{} substantially raises the bar for defending against input-only attacks in MoE LLMs.

\para{Contributions.}
Our key contributions are outlined below:

\noindent$\bullet$\textit{~New threat insight.}
We identify MoE routing as a practical attack surface for remotely hosted LLM services. Unlike prior routing-based attacks that require direct model modification, we show that routing behavior can be exploited through input-only manipulation by optimizing adversarial prompts on open-source surrogate MoE models and transferring them to same-family public API services.

\noindent$\bullet$\textit{~Input-only routing attack.}
We propose \ourmethod{}, an input-only attack framework that jointly targets routing behavior and expert functionality. \ourmethod{} identifies weakly aligned experts that are willing to produce target harmful content, steers routing away from strongly aligned experts, biases routing toward highly capable general-purpose experts to preserve output quality, and uses a two-phase optimization strategy to stabilize the interaction between routing control and output generation.

\noindent$\bullet$\textit{~Extensive evaluation.}
We evaluate \ourmethod{} across multiple MoE LLM services, datasets, and attack settings. Results show that routing-aware attacks improve ASR in transfer settings and remain effective against several defense strategies.

%% file: 2_background.tex
\section{Background}
\label{sec:back}

\subsection{Mixture of Experts}
\label{subsec:moe}

Mixture-of-Experts (MoE)~\cite{jacobs1991adaptive} has emerged as a dominant paradigm for scaling large language models~\cite{cai2025survey, shazeer2017outrageously, du2022glam}. Unlike dense models, where all parameters are activated for every input, MoE adopts \emph{sparse conditional computation}, where only a subset of experts is activated per token. This design enables significantly larger model capacity under the same computational budget (e.g., FLOPs), and has been widely adopted in modern systems such as GPT-series~\cite{agarwal2025gpt} and DeepSeek~\cite{liu2024deepseek}.

At a high level, an MoE layer consists of a set of experts $\{f_i\}_{i=1}^E$ and a router $R(\cdot)$ that selects a subset of experts for each input. Given an input $x$, the model output can be written as:
\begin{equation}
f(x) = \sum_{i \in \text{TopK}(R(x))} p_i \, f_i(x),
\end{equation}
where $\text{TopK}(\cdot)$ selects the top-$k$ experts based on routing scores, and the routing weights are given by:
\begin{equation}
p_i = \text{softmax}(z_i).
\end{equation}

The router is typically implemented as a lightweight network (e.g., a linear projection) that produces logits $z_i$ for each expert. These logits determine which experts are activated and how much they contribute to the final output.

In addition to sparse experts, many modern MoE architectures introduce \emph{shared experts}~\cite{liu2024deepseek}, which are always activated regardless of routing decisions. Shared experts serve as a dense backbone that captures general-purpose knowledge, while sparse experts specialize in specific patterns. This hybrid design can be viewed as an intermediate form between fully dense and fully sparse models, improving both stability and performance.

\subsection{Jailbreak Attacks}
\label{subsec:jailbreak}

The goal of jailbreak attacks~\cite{qi2023fine, shen2024anything, souly2024strongreject, zou2023universal, zhou2024large} is to elicit harmful or policy-violating outputs from aligned language models. For example, an attacker may attempt to bypass safety mechanisms and induce the model to generate instructions for illegal activities or produce misleading information. In practice, this has led to an iterative arms race between attacks and defenses.

Existing jailbreak attacks can be broadly categorized into prompt-engineering-based methods and optimization-based methods.

\textit{Prompt-engineering-based attacks}~\cite{shen2024anything, zhou2024large, wei2023jailbroken} aim to design adversarial prompts that exploit weaknesses in alignment behavior. These methods typically rely on semantic tricks such as role-playing, hypothetical scenarios, or instruction obfuscation to bypass safety filters, without modifying the model or performing explicit optimization.

\textit{Optimization-based attacks}~\cite{zou2023universal} formulate jailbreak as a search or optimization problem, aiming to maximize the likelihood of unsafe outputs. Let $x$ denote the input prompt and $y$ denote a target harmful response. These methods seek:
\begin{equation}
\max_{x} \ \log P(y \mid x),
\end{equation}
subject to certain constraints on the input (e.g., token budget or perturbation structure).

A representative example is the Greedy Coordinate Gradient (GCG) attack~\cite{zou2023universal}, which iteratively updates the input tokens to increase the probability of a target unsafe response. At each step, GCG selects a token position and replaces it with a candidate that maximizes the objective, guided by gradient information. Formally, given an input sequence $x = (x_1, \dots, x_n)$, GCG performs coordinate-wise updates:
\begin{equation}
x_j \leftarrow \arg\max_{w \in \mathcal{V}} \ \log P(y \mid x_{1}, \dots, x_{j-1}, w, x_{j+1}, \dots, x_n),
\end{equation}
where $\mathcal{V}$ denotes the vocabulary. This process is repeated until the model produces the desired unsafe output.

On the defense side, existing approaches can be divided into training-time alignment and deployment-time safeguards.

\textit{Training-time alignment} methods, such as reinforcement learning from human feedback (RLHF)~\cite{ouyang2022training} and direct preference optimization (DPO)~\cite{rafailov2023direct}, aim to reduce harmful outputs by incorporating safety preferences into the training objective.

\textit{Deployment-time defenses} operate at inference time~\cite{dong2024building, inan2023llama}, including content filtering and external guardrails that monitor and block unsafe inputs or outputs.

Despite these efforts, jailbreak attacks remain highly effective, especially against modern large language models. This challenge becomes even more pronounced in Mixture-of-Experts architectures, where the routing mechanism introduces additional attack surfaces, as discussed in the next subsection.

\para{Jailbreak attacks against mixture of experts.}
MoE architectures introduce a new attack surface: the routing system. Unlike dense models, where all parameters are uniformly involved, MoE models rely on dynamic expert selection, making their behavior dependent on routing decisions.

Recent attacks specifically targeting MoE systems primarily operate at the routing level by altering which experts are selected during inference~\cite{jiang2026sparse, fayyaz2025steering, wu2025gatebreaker, lai2025safex}. Formally, these methods typically operate by modifying the routing distribution:
\begin{equation}
\tilde{p}_i \propto \exp(z_i + \delta_i),
\end{equation}
where $\delta_i$ represents an adversarial perturbation applied to the routing logits.

Different works instantiate $\delta_i$ in different ways. 
Some works apply hard masking, i.e., $\delta_i \in \{0, -\infty\}$, which effectively removes selected experts from consideration~\cite{jiang2026sparse, lai2025safex}. 
In contrast, other works adopt a softer strategy, where $\delta_i$ takes finite values to reweight expert contributions~\cite{fayyaz2025steering}.

Beyond routing-level manipulation, some works also intervene in the internal computation of experts~\cite{wu2025gatebreaker}. 
Specifically, these methods apply neuron-level masking within experts, resulting in:
\begin{equation}
f_i(x) \rightarrow \tilde{f}_i(x),
\end{equation}
which further amplifies attack effectiveness by directly altering expert functionality.

Overall, existing works typically assume that the attacker can fully control the model during inference, with access to internal components such as routing logits $z_i$ and expert functions $f_i(\cdot)$, and the ability to modify both the routing process and the internal computation of experts.

\subsection{Threat Model}
\label{subsec:threat}

In contrast to prior routing-based attacks~\cite{jiang2026sparse, fayyaz2025steering, wu2025gatebreaker, lai2025safex}, we study an \emph{input-only attack} setting against MoE LLMs, where the attacker manipulates only the input and cannot modify the model. Specifically, the attacker cannot intervene in routing, alter routing logits, mask experts, or change expert parameters. Instead, the attack exploits MoE routing indirectly by crafting inputs that steer the model toward routing patterns associated with harmful compliance and reduced refusal.
This setting differs from generic jailbreak attacks, which primarily target output-level safety alignment. Our setting explicitly leverages MoE-specific routing dynamics to induce stronger unsafe behaviors by changing which experts are activated. To the best of our knowledge, this is the first work to study input-only attacks that exploit MoE routing in remotely hosted LLM services.

\para{Attacker's Goal.}
Given a target MoE LLM service and a harmful query, the attacker aims to elicit a policy-violating response using only input manipulation. The goal is to make the model follow the harmful intent rather than refuse or provide generic safety-oriented content. The attack succeeds only if the output is both non-refusal and sufficiently informative for the target harmful query.

\para{Attacker's Knowledge.}
We assume the attacker has access to an open-source surrogate MoE model and targets a remotely hosted MoE-based public API service. The surrogate may be a smaller model from the same family as the target, or a closely related open-source model with similar routing architecture and training characteristics. The attacker optimizes adversarial prompts on the surrogate in a white-box setting and transfers them to the target API.
This assumption is realistic: many commercial LLM services are built within a small number of model families, while related open-source models are increasingly available. This creates a practical surrogate-to-target setting for optimizing prompts locally and deploying them against same-family public API services. Prior work also suggests that model fingerprinting can help infer the underlying model family of a target service, further facilitating surrogate selection~\cite{chen2022teacher, shao2025sok}.

\para{Attacker's Capability.}
During attack construction, the attacker has white-box access to the open-source surrogate model, including routing behavior, expert activations, and other internal states needed for optimization. At deployment time, the attacker has only query access to the target service: they can submit arbitrary prompts and observe outputs, but cannot access routing logits, selected experts, or per-layer expert activations. The attacker also cannot modify routing decisions, mask experts, or change expert parameters. Thus, the attack on the target service must be realized purely through carefully crafted inputs.

%% file: 3_method.tex
\section{Input-only Attacks against MoE LLMs}
\label{sec:intuition}

\subsection{Motivation}

Prior works on attacking MoE LLMs typically rely on strong assumptions that allow the attacker to directly modify the model, such as altering routing decisions or expert computations~\cite{jiang2026sparse, fayyaz2025steering, wu2025gatebreaker, lai2025safex}. While effective, these attacks apply primarily to \emph{locally deployable models}, where the attacker has sufficient control over the model internals. They do not naturally extend to \emph{public API services}, where the model is remotely hosted, accessible only through input queries, and cannot be modified by the attacker.

This limitation is important because public LLM services are often the most security-critical and practically relevant targets. Compared with locally deployable open-source models, these services may incorporate proprietary model designs, larger-scale model variants, and non-public training or fine-tuning data. Proprietary designs may include undisclosed architectural choices, routing strategies, optimization recipes, and safety-training pipelines that are not available in open-source releases. Larger-scale models often provide higher model capacity and can support stronger reasoning, coding, and instruction-following abilities. Non-public training or fine-tuning data can further introduce domain-specific knowledge, service-specific behaviors, and capabilities that are difficult to reproduce in local open-source models.

This capability gap remains visible in today's LLM landscape. Despite rapid progress in open-source models, public benchmarks and leaderboards continue to show that frontier closed-source services often outperform open models on challenging capability evaluations. Therefore, attacks confined to locally controlled models cannot directly assess whether the routing mechanism of these deployed services can be exploited, nor can they expose the risks associated with service-specific capabilities and knowledge.

These limitations motivate us to study \emph{input-only attacks} against MoE LLMs. By restricting the attack to input-level manipulation, this setting naturally applies to remotely hosted public API services, where the attacker cannot inspect or modify the target model. In this work, the attacker optimizes adversarial inputs on an open-source surrogate MoE model and transfers the resulting prompts to public API services within the same model family. 
This surrogate-to-service setting allows us to examine whether MoE routing can be exploited in realistic deployments, thereby extending the routing-based attack surface beyond locally deployable models to remotely hosted LLM services.

\subsection{Problem Formulation}
\label{subsec:Problem Formulation}

Prior work has shown that safety alignment in MoE LLMs is not uniformly distributed across all experts, but is instead unevenly reflected in different expert subsets~\cite{jiang2026sparse, fayyaz2025steering, wu2025gatebreaker, lai2025safex}. Some experts are more strongly associated with refusal-style continuations, while others are more likely to support harmful compliance. As a result, the routing outcome directly affects attack success, since different activated expert subsets can lead to different levels of refusal, compliance, and output quality.

However, routing alone does not determine attack success. Even if the input is routed away from strongly aligned experts, the selected experts may still retain residual safety alignment and refuse the harmful query~\cite{wu2025gatebreaker}. Conversely, experts that are more willing to produce unsafe continuations may not necessarily generate high-quality or informative responses. Therefore, an effective input-only attack against MoE LLMs must satisfy two coupled requirements: it must steer the routing mechanism toward an attack-favorable routing pattern, and it must induce the resulting activated experts to generate a policy-violating response with sufficient quality.

This leads to a natural formulation of input-only attacks against MoE LLMs with two coupled objectives:

$\bullet$\textbf{~Routing steering:} influencing the routing outcome so that the input is assigned to an attack-favorable expert subset, e.g., one that reduces reliance on strongly aligned experts while favoring experts more likely to support harmful compliance and useful generation.

$\bullet$\textbf{~Output attack:} inducing the activated experts to generate a non-refusal, policy-violating, and sufficiently informative response to the harmful query.

Formally, let $x$ denote a harmful query and $\tilde{x}$ denote the adversarial prompt constructed by the attacker, e.g., by adding a prefix to $x$. Let $R(\cdot)$ denote the routing function and let
\begin{equation}
    G(\tilde{x}) = \mathrm{TopK}(R(\tilde{x}))
\end{equation}
denote the selected expert subset for $\tilde{x}$. The attacker aims to optimize $\tilde{x}$ such that:
\begin{equation}
    \text{(i)} \quad G(\tilde{x}) \in \mathcal{G}_{\mathrm{adv}},
\end{equation}
\begin{equation}
    \text{(ii)} \quad y^*(\tilde{x}) \in \mathcal{Y}_{\mathrm{harmful}}
    \quad \text{and} \quad
    Q(y^*(\tilde{x}), x) \geq \tau,
\end{equation}
where $\mathcal{G}_{\mathrm{adv}}$ denotes the set of attack-favorable routing patterns, $\mathcal{Y}_{\mathrm{harmful}}$ denotes the set of policy-violating outputs, $y^*(\tilde{x})$ denotes the model response to $\tilde{x}$, and $Q(\cdot, \cdot)$ measures whether the response is sufficiently informative for the target harmful query.

The first objective steers the routing mechanism toward expert subsets that are more favorable for unsafe generation, while the second objective ensures that the final response is not merely a non-refusal, but also contains useful harmful content. Notably, given a fixed routing outcome, the MoE model behaves like a fixed subnetwork over the selected experts. In this case, existing optimization-based jailbreak attacks, such as GCG~\cite{zou2023universal}, can be applied to optimize the output objective. The key difficulty in MoE LLMs is that the routing outcome is itself input-dependent and can change during prompt optimization.

This formulation highlights a fundamental difference from jailbreak attacks in dense models. In dense models, the attack mainly targets the output distribution. In contrast, input-only attacks against MoE LLMs involve a \emph{multi-objective optimization problem} over both routing behavior and output generation. These two objectives are coupled: optimizing the prompt for harmful outputs may change the routing pattern, while enforcing a desired routing pattern may restrict the output attack. This coupling is the central challenge addressed by \ourmethod{}.

\subsection{Challenges}
\label{subsec:challenge}

As discussed above, input-only attacks against MoE LLMs can be formulated as a \emph{multi-objective optimization problem}, where the attacker must simultaneously influence routing behavior and induce policy-violating outputs. This introduces several unique challenges.

\para{(i) Limited controllability over routing.}
Unlike prior routing-based attacks that assume direct access to and modification of the model, such as manipulating routing logits or expert computations, input-only attacks can only influence routing indirectly through prompt perturbations. As a result, the attacker can only approximate a desired routing pattern rather than explicitly enforce it. This limitation becomes especially pronounced when the target routing behavior deviates substantially from the model's default routing behavior. In such cases, input manipulation alone may fail to reliably induce the intended expert selection. Therefore, choosing an attainable routing objective is crucial: an overly rigid objective, such as completely avoiding a fixed set of strongly aligned experts, may be infeasible under input-only constraints.

\para{(ii) Optimization difficulty under coupled objectives.}
Routing control and output generation are inherently coupled. Input changes that improve the output attack may alter the routing pattern, while constraints imposed to preserve a desired routing pattern may hinder the generation of policy-violating responses. Therefore, progress on one objective does not necessarily translate into progress on the other, and may even degrade it. This coupling creates an unstable optimization landscape, where the attack can easily converge to suboptimal solutions. An effective input-only attack must therefore coordinate routing control and output generation without either objective to dominate or destabilize the other.

\para{(iii) Trade-off between safety bypass and output quality.}
Even when the attacker successfully reduces refusal behavior, the resulting output may still be of low quality. Routing the input away from strongly aligned experts may help bypass refusal-style continuations, but it does not guarantee that the selected experts are capable of producing a coherent and sufficiently informative response to the target query. For example, the input may be routed to experts that are weakly aligned but poorly suited for the task, leading to outputs that are vague, inconsistent, or unhelpful. A successful attack must therefore not only induce non-refusal behavior, but also preserve enough generation quality for the response to be useful. This creates a challenge specific to MoE LLMs: the attack must exploit routing for safety bypass while still maintaining the functional capability of the activated experts.

\subsection{Key Insight}

\para{(i) From routing avoidance to routing shaping.}
Prior work has shown that safety alignment in MoE LLMs is not uniformly distributed across all experts, but is instead reflected unevenly across different expert subsets. A natural input-only extension of prior routing-based attacks is therefore to steer routing away from strongly aligned experts that are associated with refusal-style continuations.
For example, following the profiling strategy used in GateBreaker~\cite{wu2025gatebreaker}, one can identify strongly aligned experts by probing the model with harmful queries and observing which experts are activated when the model produces refusal-style continuations. Given a set of harmful queries $\mathcal{D}_{\text{harm}}$, we estimate how frequently each expert $i$ is activated as:
\begin{equation}
\label{eq:gatebreaker}
U_i^{\text{harm}} =
\mathbb{E}_{x \in \mathcal{D}_{\text{harm}}}
\left[
\mathbb{I}\left(i \in \mathrm{TopK}(R(x))\right)
\right],
\end{equation}
where $R(x)$ denotes the router, $\mathrm{TopK}(R(x))$ denotes the selected experts, and $\mathbb{I}(\cdot)$ is the indicator function. Experts with high $U_i^{\text{harm}}$ are treated as strongly aligned experts, since they are frequently activated when the model processes harmful queries and produces refusal-style behavior.

A straightforward strategy is then to construct adversarial prompts that reduce the routing probability assigned to these experts:
\begin{equation}
\min_{\tilde{x}} \sum_{i \in \mathcal{S}} p_i(\tilde{x}),
\end{equation}
where $\mathcal{S}$ denotes the set of strongly aligned experts and $p_i(\tilde{x})$ denotes the routing probability of expert $i$ for adversarial input $\tilde{x}$.

However, as discussed in \autoref{subsec:challenge}, this routing-avoidance objective is often too rigid under input-only constraints. Since routing can only be influenced indirectly through prompt perturbations, the attacker may fail to reliably steer the input away from a fixed set of experts, especially when the desired routing pattern deviates substantially from the model's default behavior. More importantly, avoiding strongly aligned experts only reduces reliance on refusal-related routing patterns; it does not ensure that the activated experts are willing to produce the target harmful content.

Our key insight is therefore to move from \emph{routing avoidance} to \emph{routing shaping}. Instead of only suppressing strongly aligned experts, we explicitly identify weakly aligned experts that are willing to produce target harmful content. To this end, we construct two datasets that provide contrasting routing signals:

$\bullet~\mathcal{D}_{\text{harm}}$: a set of harmful queries used to identify routing patterns associated with strongly aligned behavior, such as refusal-style continuations (e.g., ``I can't help with that...'').

$\bullet~\mathcal{D}_{\text{comp}}$: a set of harmful queries paired with unsafe continuations. Here, ``comp'' denotes harmful compliance. Specifically, each sample concatenates a harmful query with an unsafe continuation that follows the harmful intent, and incorporates basic jailbreak strategies such as semantic paraphrasing, adversarial perturbations, and context reframing to increase diversity. We feed the resulting sequence into the model and analyze the expert activations to identify experts associated with harmful compliance, i.e., weakly aligned experts.

These two datasets capture different expert behaviors. $\mathcal{D}_{\text{harm}}$ highlights experts associated with strong safety alignment and refusal, while $\mathcal{D}_{\text{comp}}$ highlights experts that are more willing to continue with harmful content. We therefore use these signals contrastively: the attack should steer routing away from strongly aligned experts and toward weakly aligned experts.

Formally, we define:
\begin{equation}
U_i^{\text{comp}} =
\mathbb{E}_{x \in \mathcal{D}_{\text{comp}}}
\left[
\mathbb{I}\left(i \in \mathrm{TopK}(R(x))\right)
\right].
\end{equation}
We then construct a weighted contrast:
\begin{equation}
\label{eq:ui}
\Delta U_i =
\lambda_1 U_i^{\text{harm}}
-
\lambda_2 U_i^{\text{comp}},
\end{equation}
where $\lambda_1, \lambda_2 > 0$ control the relative strength of suppressing strongly aligned experts and promoting weakly aligned experts. Intuitively, experts with large positive $\Delta U_i$ are more associated with refusal-style behavior, while experts with smaller or negative $\Delta U_i$ are more associated with harmful compliance. During routing optimization, minimizing a routing score weighted by $\Delta U_i$ discourages activation of strongly aligned experts while encouraging activation of weakly aligned experts.

This formulation provides a more attainable objective under input-only constraints. Rather than approximating a hard expert mask, the attack follows a directional routing signal that both suppresses refusal-related patterns and promotes alternative patterns associated with harmful compliance.

\para{(ii) Preserving output quality with general-purpose experts.}
Routing the input toward weakly aligned experts improves the chance of reducing refusal, but willingness to answer does not imply the ability to answer well. In practice, some experts may be weakly aligned yet poorly suited for the target task, leading to responses that are vague, incoherent, or uninformative. Therefore, an effective input-only attack must not only reduce refusal behavior, but also preserve sufficient output quality.

Our observation is that although MoE models distribute computation across different experts, some experts are consistently activated by diverse benign question-answering tasks and are associated with stronger general generation capability (See \autoref{subsec:mask}). We refer to them as \emph{highly capable general-purpose experts}. These experts are useful for maintaining output quality because they are not tied to a single narrow behavior, but instead contribute to coherent and informative responses across a broad range of benign tasks.
To identify such experts, we construct an additional dataset $\mathcal{D}_{\text{benign}}$ consisting of benign question-answering queries, and estimate:
\begin{equation}
U_i^{\text{benign}} =
\mathbb{E}_{x \in \mathcal{D}_{\text{benign}}}
\left[
\mathbb{I}\left(i \in \mathrm{TopK}(R(x))\right)
\right].
\end{equation}
Experts with high $U_i^{\text{benign}}$ are treated as highly capable general-purpose experts.

We then incorporate this utility signal into the routing objective:
\begin{equation}
\label{eq:ui_benign}
\Delta U_i =
\lambda_1 U_i^{\text{harm}}
-
\lambda_2 U_i^{\text{comp}}
-
\lambda_3 U_i^{\text{benign}},
\end{equation}
where $\lambda_3 > 0$ controls the strength of the utility anchor. Under this sign convention, minimizing the routing score induced by $\Delta U_i$ suppresses strongly aligned experts, promotes weakly aligned experts, and further biases routing toward highly capable general-purpose experts. As a result, the attack can reduce refusal behavior while maintaining a basic level of response quality.

\para{(iii) Decoupling routing control and output generation.}
Recall from \autoref{subsec:Problem Formulation} that input-only attacks against MoE LLMs involve two coupled objectives: routing control and output generation. A straightforward approach is to combine them into a single loss. However, this often yields unstable optimization, because input changes that improve harmful generation may alter routing decisions, while constraints that preserve routing may hinder the output attack.
Our key insight is to decouple these objectives through a two-phase optimization strategy. In the first phase, we focus solely on routing control, optimizing the routing objective derived from \autoref{eq:ui_benign}. This allows the adversarial input to reach an attack-favorable routing pattern without interference from output-related objectives.
Once this routing pattern is established, the problem effectively reduces to attacking a more stable set of activated experts. In the second phase, we optimize the output objective to induce harmful generation while retaining a routing term to preserve routing stability:
\begin{equation}
\mathcal{L}
=
\alpha \mathcal{L}_{\text{out}}
+
\beta \mathcal{L}_{\text{route}},
\end{equation}
where $\mathcal{L}_{\text{route}}$ is derived from the routing objective in \autoref{eq:ui_benign}, and $\alpha,\beta > 0$ control the trade-off between output attack and routing stability.
This design acts as routing regularization. The first phase identifies an attainable routing region, while the second phase improves harmful generation without allowing the prompt to drift away from that region. As a result, \ourmethod{} can jointly exploit weakly aligned experts, preserve output quality through highly capable general-purpose experts, and stabilize the interaction between routing control and output generation.

\begin{figure*}[!t]
\centerline{\includegraphics[width=\linewidth]{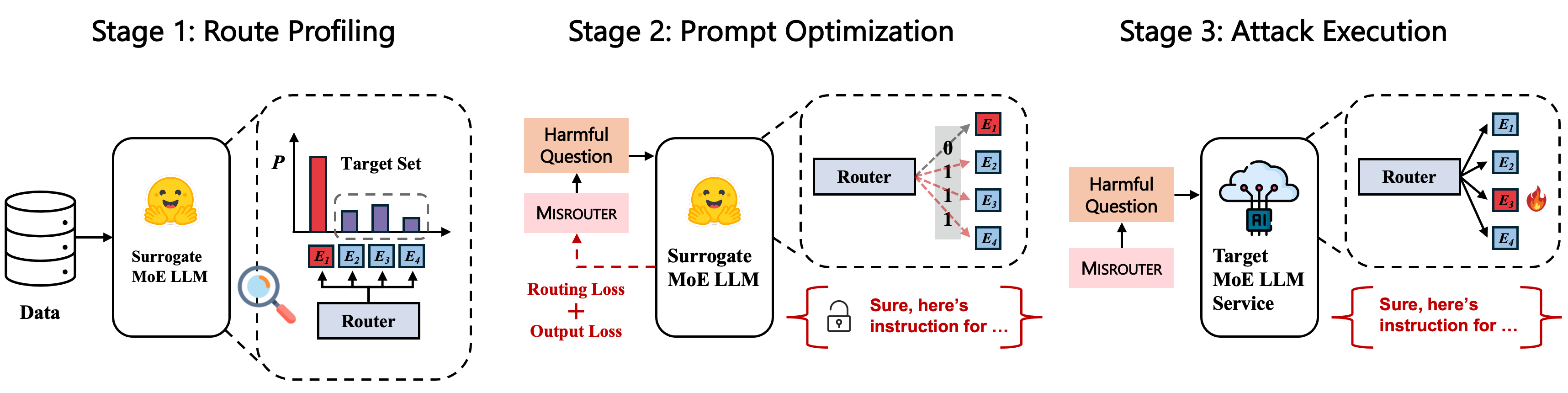}}
\caption{Overview of \ourmethod{}.}
\Description{}
\label{fig:overview}
\end{figure*}

\section{Design of \ourmethod{}}
\label{sec:ourmethod}

\subsection{Overview}

In this section, we present the design of \ourmethod{}, which constructs adversarial prompts by characterizing routing behaviors under different input distributions and optimizing inputs to induce both attack-favorable routing and harmful outputs. 
\autoref{fig:overview} illustrates the overview of \ourmethod{} with four stages.

\paragraph{Stage 0: Data construction.}
We first construct three datasets that capture different aspects of model behavior: a harmful query set $\mathcal{D}_{\text{harm}}$, a harmful-compliance set $\mathcal{D}_{\text{comp}}$, and a benign query set $\mathcal{D}_{\text{benign}}$. 
The harmful set consists of harmful queries in their original form and is used to identify routing patterns associated with strongly aligned behavior, such as refusal-style continuations. 
The harmful-compliance set pairs harmful queries with unsafe continuations, e.g., ``Sure, here are instructions for...'', and is used to identify weakly aligned experts that are willing to produce target harmful content. 
Here, ``comp'' denotes harmful compliance.
The benign set contains regular question-answering queries and is used to identify highly capable general-purpose experts that contribute to coherent and informative responses.
Together, these datasets provide complementary signals for characterizing refusal behavior, harmful compliance, and generation utility.

\paragraph{Stage 1: Route profiling.}
Using these datasets, we estimate routing statistics for each expert based on their activation frequencies on an open-source surrogate MoE model. 
We then construct a weighted routing signal that suppresses strongly aligned experts, promotes weakly aligned experts, and biases routing toward highly capable general-purpose experts.

To improve robustness beyond direct empirical estimation, we adopt a resampling strategy. Specifically, we repeatedly sample subsets from each dataset, recompute routing statistics, and aggregate the results to identify consistently activated experts. This reduces sensitivity to particular input samples and yields a more stable estimate of the desired routing preference.

\paragraph{Stage 2: Prompt optimization.}
We then optimize adversarial prompts on the surrogate model to realize the desired routing behavior and induce harmful outputs. 
The optimization follows a two-phase procedure: we first optimize the prompt to reach an attack-favorable routing pattern, and then refine the prompt to improve the output objective while preserving routing stability.

\paragraph{Stage 3: Attack execution.}
Once constructed, the optimized prompts are directly applied to the target public API service via standard query access. 
This attack relies on surrogate-to-target transferability: routing behaviors and expert specializations learned from an open-source surrogate model can provide useful signals for attacking same-family public API services. 
As a result, \ourmethod{} can exploit MoE routing in remotely hosted services without requiring access to their internal states.

Intuitively, this design can yield better black-box transferability than standard output-level optimization, such as GCG~\cite{zou2023universal}. GCG directly optimizes against the full output distribution of the surrogate model, which depends on the combined behavior of many activated experts and is therefore highly model-specific. In contrast, \ourmethod{} targets the routing mechanism, a comparatively simpler and lower-dimensional decision component. As a result, routing-aware prompts are more likely to preserve useful attack signals when transferred from an open-source surrogate model to a same-family public API service.

In the following subsections, we formally describe the two main components: route profiling and prompt optimization.

\subsection{Route Profiling}

In this section, we formalize the route profiling procedure used to construct the routing preference for \ourmethod{}. 
The inputs are three datasets $\mathcal{D}_{\text{harm}}$, $\mathcal{D}_{\text{comp}}$, and $\mathcal{D}_{\text{benign}}$, together with an open-source surrogate MoE model. 
The output is a routing score vector $\Delta U$, where each entry assigns a signed preference to an expert. 

Let the MoE model consist of $N$ experts indexed by $i \in \{1, \dots, N\}$. 
Given an input $x$, the router produces logits $z_i(x)$ and routing probabilities $p_i(x) \propto \exp(z_i(x))$. 
We denote by $\mathrm{TopK}(R(x))$ the set of $K$ selected experts for input $x$. 
For simplicity, $p_i(x)$ and $\mathrm{TopK}(R(x))$ can be understood as routing statistics aggregated over the relevant layers and token positions.

\paragraph{Resampling protocol.}
Let $\mathcal{D} \in \{\mathcal{D}_{\text{harm}}, \mathcal{D}_{\text{comp}}, \mathcal{D}_{\text{benign}}\}$ denote one of the profiling datasets. 
We construct $S$ resampled subsets $\{\mathcal{D}^{(s)}\}_{s=1}^S$, where each subset $\mathcal{D}^{(s)}$ contains $m$ inputs sampled from $\mathcal{D}$, with or without replacement.

For each subset $\mathcal{D}^{(s)}$, we estimate the activation frequency of expert $i$ as:
\begin{equation}
U_i^{(s)}(\mathcal{D}) =
\frac{1}{m}
\sum_{x \in \mathcal{D}^{(s)}}
\mathbb{I}\big(i \in \mathrm{TopK}(R(x))\big),
\end{equation}
where $\mathbb{I}(\cdot)$ is the indicator function. 
The parameter $S$ controls the number of resampled subsets and thus the stability of the estimation, while $m$ controls the size of each subset and thus the variance of the frequency estimate.

We then aggregate across resamples:
\begin{equation}
\bar{U}_i(\mathcal{D}) =
\frac{1}{S}
\sum_{s=1}^{S}
U_i^{(s)}(\mathcal{D}).
\end{equation}

\paragraph{Weighted routing contrast.}
We compute the routing score for each expert $i$ as:
\begin{equation}
\label{eq:route-score}
\Delta U_i =
\lambda_1 \bar{U}_i(\mathcal{D}_{\text{harm}})
-
\lambda_2 \bar{U}_i(\mathcal{D}_{\text{comp}})
-
\lambda_3 \bar{U}_i(\mathcal{D}_{\text{benign}}),
\end{equation}
where $\lambda_1, \lambda_2, \lambda_3 > 0$ control the relative importance of suppressing strongly aligned experts, promoting weakly aligned experts, and preserving generation utility, respectively.

This sign convention is important. 
Experts frequently activated by $\mathcal{D}_{\text{harm}}$ receive larger positive scores, since they are more associated with strongly aligned behavior and refusal-style continuations. 
Experts frequently activated by $\mathcal{D}_{\text{comp}}$ receive lower scores, since they are more associated with harmful compliance. 
Experts frequently activated by $\mathcal{D}_{\text{benign}}$ also receive lower scores, since they are more likely to be highly capable general-purpose experts that support coherent and informative generation. 
Therefore, minimizing a routing objective weighted by $\Delta U_i$ discourages routing toward strongly aligned experts while encouraging routing toward weakly aligned and highly capable general-purpose experts.

Optionally, one can derive two expert subsets from the score vector:
\begin{equation}
\mathcal{E}_{\text{sup}} = \operatorname{TopK}\big(\{\Delta U_i\}_{i=1}^{N}, k_{\text{sup}}\big),
\end{equation}
\begin{equation}
\mathcal{E}_{\text{pro}} = \operatorname{BottomK}\big(\{\Delta U_i\}_{i=1}^{N}, k_{\text{pro}}\big),
\end{equation}
where $\mathcal{E}_{\text{sup}}$ denotes experts to be suppressed and $\mathcal{E}_{\text{pro}}$ denotes experts to be promoted. 
In our optimization, however, we use the full score vector $\Delta U$ by default, as it provides a smoother routing signal than a hard expert mask.

\para{The algorithm.}
We summarize the route profiling procedure in \autoref{alg:route-profiling} in \autoref{subsec:algorithm}.

\subsection{Prompt Optimization}

In this section, we formalize the prompt optimization procedure used to construct adversarial inputs.

\para{Problem setup.}
Given a harmful query $x$, the routing score vector $\Delta U$, and a surrogate MoE model with router $R$, the goal is to construct an adversarial prefix $\delta = (t_1, \dots, t_L)$ of fixed length $L$, such that the modified input $\tilde{x} = \delta \oplus x$ induces an attack-favorable routing pattern and produces a harmful output.

\paragraph{Phase I: routing steering.}
In the first phase, we optimize the prefix $\delta$ only for routing control. 
Instead of suppressing a fixed set of experts, we optimize the prompt using the signed routing preference $\Delta U$. 
The routing objective is defined as:
\begin{equation}
\label{eq:l_route}
\mathcal{L}_{\text{route}}(\delta; x)
=
\sum_{i=1}^{N}
\Delta U_i \, p_i(\tilde{x}),
\end{equation}
where $p_i(\tilde{x})$ denotes the routing probability of expert $i$ for input $\tilde{x} = \delta \oplus x$.

Minimizing \autoref{eq:l_route} discourages routing toward experts with large positive $\Delta U_i$, which are more associated with strongly aligned behavior, while encouraging routing toward experts with lower or negative $\Delta U_i$, which are more associated with harmful compliance and general-purpose generation capability.

We perform $T_1$ iterations of gradient-based token optimization over $\delta$. 
At each iteration, we compute the gradient of $\mathcal{L}_{\text{route}}$ with respect to the input token embeddings, and update the tokens in $\delta$ to reduce the routing loss.

\paragraph{Phase II: joint optimization.}
In the second phase, we optimize $\delta$ to induce harmful outputs while preserving routing stability.

Let $y^{*} = (y_1^{*}, \dots, y_T^{*})$ denote a target unsafe continuation. Following GCG-style optimization, we define the output loss:
\begin{equation}
\mathcal{L}_{\text{out}}(\delta; x)
=
-
\sum_{t=1}^{T'}
\log P\big(y_t^{*} \mid \tilde{x}, y_{<t}^{*}\big),
\end{equation}
where $T'$ is the number of target tokens used for optimization.

The joint objective is:
\begin{equation}
\label{eq:joint-loss}
\mathcal{L}(\delta; x)
=
\alpha \mathcal{L}_{\text{out}}(\delta; x)
+
\beta \mathcal{L}_{\text{route}}(\delta; x),
\end{equation}
where $\alpha, \beta > 0$ balance output optimization and routing stability.

This formulation preserves the routing pattern established in Phase I while improving the likelihood of producing the target harmful output. 
The routing term acts as a regularizer that prevents the prompt from drifting back toward strongly aligned experts during output optimization.

We perform $T_2$ iterations of optimization in this phase.

\paragraph{Transfer to target service.}
After optimization, the adversarial prefix $\delta$ is concatenated with the original query $x$ and directly applied to the target public API service. 
The target service is accessed only through standard input queries, and the attacker does not observe routing logits, selected experts, or other internal states. 
Thus, the attack against the target service remains purely input-only.

\para{The algorithm.}
We summarize the prompt optimization procedure in \autoref{alg:prompt-opt} in \autoref{subsec:algorithm}.

%% file: 4_evaluation.tex
\begin{table*}[!t]
\caption{Attack success rates (ASR) and routing losses of baseline attacks and \ourmethod{} in the white-box setting.}
\resizebox{\textwidth}{!}{
\begin{tabular}{cccccccccccccc}
\toprule
\multirow{2}{*}{Dataset} & \multirow{2}{*}{Method} & \multicolumn{2}{c}{GPT-OSS} & \multicolumn{2}{c}{Qwen3} & \multicolumn{2}{c}{Phi-3.5} & \multicolumn{2}{c}{Mixtral} & \multicolumn{2}{c}{Qwen1.5} & \multicolumn{2}{c}{Deepseek} \\
\cmidrule{3-14}
 &  & \multicolumn{1}{c}{ASR} & \multicolumn{1}{c}{\begin{tabular}[c]{@{}c@{}}Routing\\ Loss\end{tabular}} & \multicolumn{1}{c}{ASR} & \multicolumn{1}{c}{\begin{tabular}[c]{@{}c@{}}Routing\\ Loss\end{tabular}} & \multicolumn{1}{c}{ASR} & \multicolumn{1}{c}{\begin{tabular}[c]{@{}c@{}}Routing\\ Loss\end{tabular}} & \multicolumn{1}{c}{ASR} & \multicolumn{1}{c}{\begin{tabular}[c]{@{}c@{}}Routing\\ Loss\end{tabular}} & \multicolumn{1}{c}{ASR} & \multicolumn{1}{c}{\begin{tabular}[c]{@{}c@{}}Routing\\ Loss\end{tabular}} & \multicolumn{1}{c}{ASR} & \multicolumn{1}{c}{\begin{tabular}[c]{@{}c@{}}Routing\\ Loss\end{tabular}} \\
\midrule
\multirow{7}{*}{AdvBench} & Direct Instruction & 0.0\% & 18.46 & 0.0\% & 13.52 & 0.0\% & 9.27 & 8.0\% & 10.06 & 0.0\% & 3.95 & 10.0\% & 5.03 \\
 & GCG~\cite{zou2023universal} & 0.0\% & 18.48 & 0.0\% & 12.03 & 44.0\% & 8.88 & 18.0\% & 9.59 & 76.0\% & 3.91 & 48.0\% & 4.78 \\
 & FFA~\cite{zhou2024large} & 0.0\% & 18.45 & 0.0\% & 12.00 & 0.0\% & 9.18 & 66.0\% & 8.53 & 52.0\% & 3.70 & 98.0\% & 4.34 \\
 & Jailbroken~\cite{wei2023jailbroken} & 0.0\% & 18.51 & 0.0\% & 9.45 & 32.0\% & 8.28 & 78.0\% & 8.00 & 40.0\% & 3.47 & 20.0\% & 3.71  \\
 & WildPrompt~\cite{shen2024anything} & 2.0\% & 18.45 & 2.0\% & 10.31 & 10.0\% & 8.73 & 72.0\% & 8.06 & 64.0\% & 3.47 & 74.0\% & 3.80  \\
\cmidrule{2-14}
 & \textbf{\ourmethod{}} & \textbf{24.0}\% & 18.34 & 2.0\% & 11.98 & 52.0\% & 8.55 & 22.0\% & 9.19 & \textbf{96.0\%} & 3.78 & 62.0\% & 4.53 \\
 & \textbf{\ourmethod{}+FFA} & 10.0\% & 18.33 & \textbf{14.0\%} & 10.69 & \textbf{72.0\%} & 8.88 & \textbf{82.0\%} & 8.39 & 80\%  & 3.58 & \textbf{100.0\%} & 4.22 \\
\midrule
\multirow{7}{*}{StrongREJECT} & Direct Instruction & 0.0\% & 18.44 & 0.0\% & 12.84 & 0.0\% & 9.29 & 1.7\% & 9.25 & 5.0\% & 3.93 & 13.3\% & 4.56 \\
 & GCG~\cite{zou2023universal} & 0.0\% & 18.46 & 0.0\% & 11.92 & 46.7\% & 8.80 & 10.0\% & 9.04 & 63.3\% & 3.91 & 53.3\% & 4.40 \\
 & FFA~\cite{zhou2024large} & 0.0\% & 18.46 & 0.0\% & 11.92 & 3.3\% & 9.15 & 63.3\% & 8.49 & 50.0\% & 3.74 & 81.7\% & 4.29 \\
 & Jailbroken~\cite{wei2023jailbroken} & 3.3\% & 18.51 & 1.7\% & 9.72 & 23.3\% & 8.36 & 60.0\% & 8.07 & 51.7\% & 3.47 & 35.0\% & 3.74  \\
 & WildPrompt~\cite{shen2024anything} & 0.0\% & 18.45 & 0.0\% & 10.48 & 10.0\% & 8.74 & 58.3\% & 7.98 & 70.0\% & 3.45 & 66.7\% & 3.76  \\
\cmidrule{2-14}
 & \textbf{\ourmethod{}} & \textbf{11.7\%} & 18.36 & 5.0\% & 11.75 & 50.0\% & 8.74 & 16.7\% & 8.89 & 80.0\% & 3.86 & 63.3\% & 4.13 \\
 & \textbf{\ourmethod{}+FFA} & 6.7\% & 18.32 & \textbf{13.3\%} & 10.71 & \textbf{68.3\%} & 8.89 & \textbf{70.0\%} & 8.38 & \textbf{91.7\%} & 3.68 & \textbf{93.3\%} & 4.15 \\
\bottomrule
\end{tabular}
}
\label{tab:main}
\end{table*}

\section{Evaluation}
\label{sec:eval}

\subsection{Experimental Setup}
\label{subsec:setup}

\para{Models.}
We evaluate \ourmethod{} on six representative MoE LLM API services from five major developers: GPT-4o mini from OpenAI, Qwen-Plus and Qwen-Turbo from Alibaba, Phi-4~\cite{abdin2024phi} from Microsoft, Mistral-Large-2512 from Mistral, and DeepSeek-Chat-V3-0324~\cite{liu2024deepseek} from DeepSeek. 
For each target service, we construct attacks on open-source, locally deployable surrogate MoE models from the same developer or model family whenever available. The surrogate models are summarized in \autoref{tab:Surrogate}, and cover diverse MoE designs, including sparse MoEs, mixture MoEs, and grouped mixture MoEs.

\para{Datasets and baselines.}
We evaluate attacks on AdvBench~\cite{zou2023universal} and StrongREJECT~\cite{souly2024strongreject}, two widely used harmful-query benchmarks for jailbreak evaluation. 
We compare \ourmethod{} with representative input-only jailbreak baselines, including GCG~\cite{zou2023universal}, FFA~\cite{zhou2024large}, Jailbroken~\cite{wei2023jailbroken}, and WildPrompt~\cite{shen2024anything}. 
These baselines include both optimization-based and prompt-engineering-based attacks, allowing us to evaluate whether exploiting MoE routing provides additional benefits beyond conventional input-only jailbreak strategies. 
We also evaluate composability by applying \ourmethod{} on top of FFA prompts, denoted as \ourmethod{}+FFA.

\para{Metrics and configuration.}
We use Attack Success Rate (ASR)~\cite{krauss2025twinbreak} as the primary metric, where a response is counted as successful if it is classified as unsafe or policy-violating by Llama-Guard-3-8B~\cite{inan2023llama}. 
For open-source surrogate models, we additionally report routing loss to measure whether the optimized prompt successfully suppresses strongly aligned experts while promoting weakly aligned and highly capable general-purpose experts.
Unless otherwise specified, we use a fixed configuration across models and datasets. 
The details of dataset construction, baseline implementation, metric definitions, and hyperparameter settings are provided in \autoref{app:setup}.

\begin{table*}[!t]
\caption{Attack success rates (ASR) of baseline attacks and \ourmethod{} in the black-box transfer setting (from open-source surrogate MoE models to commercial API services) on AdvBench~\cite{zou2023universal} and StrongREJECT~\cite{souly2024strongreject}.}
\resizebox{.85\linewidth}{!}{
\begin{tabular}{cccccccc}
\toprule
Dataset & Method & GPT-4o mini & Qwen-Plus & Phi-4 & Mistral-Large & Qwen-Turbo & DeepSeek-Chat-V3 \\
\midrule
\multirow{7}{*}{AdvBench} 
 & Direct Instruction & 0.0\% & 0.0\% & 0.0\% & 6.0\% & 2.0\% & 2.0\% \\
 & GCG~\cite{zou2023universal} & 0.0\% & 0.0\% & 0.0\% & 0.0\% & 0.0\% & 0.0\% \\
 & FFA~\cite{zhou2024large} & 2.0\% & 0.0\% & 0.0\% & 92.0\% & 20.0\% & 96.0\% \\
 & Jailbroken~\cite{wei2023jailbroken} & 12.0\% & 0.0\% & 0.0\% & 86.0\% & 20.0\% & 30.0\% \\
 & WildPrompt~\cite{shen2024anything} & 0.0\% & 4.0\% & 0.0\% & 96.0\% & 22.0\% & 32.0\% \\
\cmidrule{2-8}
 & \textbf{\ourmethod{}} & 0.0\% & 0.0\% & 6.0\% & 10.0\% & 8.0\% & 2.0\% \\
 & \textbf{\ourmethod{}+FFA} & \textbf{18.0\%} & \textbf{6.0\%} & \textbf{10.0\%} & \textbf{96.0\%} & \textbf{26.0\%} & \textbf{96.0\%} \\
\midrule
\multirow{7}{*}{StrongREJECT} 
 & Direct Instruction & 0.0\% & 0.0\% & 0.0\% & 6.7\% & 1.7\% & 0.0\% \\
 & GCG~\cite{zou2023universal} & 0.0\% & 0.0\% & 0.0\% & 0.0\% & 1.7\% & 1.7\% \\
 & FFA~\cite{zhou2024large} & 5.0\% & 1.7\% & 0.0\% & 93.3\% & 28.3\% & 83.3\% \\
 & Jailbroken~\cite{wei2023jailbroken} & \textbf{11.7\%} & 3.3\% & 3.3\% & 75.0\% & 25.0\% & 20.0\% \\
 & WildPrompt~\cite{shen2024anything} & 1.7\% & 1.7\% & 1.7\% & 91.7\% & 21.7\% & 20.0\% \\
\cmidrule{2-8}
 & \textbf{\ourmethod{}} & 1.7\% & 0.0\% & 1.7\% & 8.3\% & 5.0\% & 3.3\% \\
 & \textbf{\ourmethod{}+FFA} & 5.0\% & \textbf{8.3\%} & \textbf{3.3\%} & \textbf{93.3\%} & \textbf{31.7\%} & \textbf{85.0\%} \\
\bottomrule
\end{tabular}
}
\label{tab:black}
\end{table*}

\subsection{White-Box Results}
\label{subsec:white}

In this section, we evaluate \ourmethod{} in a setting where the surrogate model and the target model are identical. This setting is practically relevant because some hosted services expose the same open-source models through API endpoints, allowing users to access them without local computational resources. In such cases, adversarial prompts optimized on the open-source model can be directly reused against the corresponding hosted service through input-only queries.

We start with this setting to isolate the effect of routing manipulation itself and avoid confounding factors introduced by cross-model transfer. The goal is not to evaluate transferability, but to verify whether \ourmethod{} works as intended by exploiting MoE routing. We study transfer settings separately in \autoref{subsec:black} and \autoref{subsec:gray}.

\autoref{tab:main} summarizes the results. Across the six open-source MoE LLMs and two benchmark datasets described in \autoref{subsec:setup}, \ourmethod{} and \ourmethod{}+FFA consistently achieve stronger attack performance than existing input-only jailbreak baselines, including both optimization-based~\cite{zou2023universal} and prompt-engineering-based attacks~\cite{zhou2024large, wei2023jailbroken, shen2024anything}. For example, on GPT-OSS, the strongest baseline achieves only 2.0\% ASR on AdvBench and 3.3\% ASR on StrongREJECT. In comparison, \ourmethod{} improves ASR to 24.0\% and 11.7\%, corresponding to 12.0$\times$ and 3.5$\times$ improvements, respectively.

To better understand the mechanism of \ourmethod{}, we further analyze the routing loss achieved by different attacks. This metric helps decouple routing manipulation from final output generation: two methods may achieve similar ASR while relying on different routing behaviors. We find that \ourmethod{} consistently achieves lower routing loss than GCG~\cite{zou2023universal}, which is the most direct comparison since both methods are optimization-based. This result confirms that \ourmethod{} improves over GCG not merely by optimizing the output objective, but by explicitly steering routing toward attack-favorable expert patterns.

Interestingly, some prompt-engineering-based attacks occasionally achieve even lower routing loss than \ourmethod{}, despite not explicitly optimizing for routing. This suggests that certain heuristic prompt transformations can incidentally steer the model toward weakly aligned or highly capable general-purpose experts and away from strongly aligned ones. However, lower routing loss alone does not necessarily translate into higher attack success. Since these methods do not jointly optimize the output objective, they may reach favorable routing patterns but still fail to elicit harmful and sufficiently informative responses. In contrast, \ourmethod{} jointly optimizes routing behavior and model output, which explains its stronger overall ASR.

\subsection{Black-Box Results}
\label{subsec:black}

In this section, we investigate the transferability of \ourmethod{}, i.e., whether adversarial prompts optimized on open-source surrogate MoE models can be directly applied to public API services within the same model family. The surrogate models and their corresponding target MoE LLM services are summarized in \autoref{tab:Surrogate} in \autoref{subsec:model}.

\autoref{tab:black} summarizes the black-box transfer results. Overall, \ourmethod{}+FFA improves over existing input-only jailbreak baselines in several target services, especially when baseline attacks have low or moderate ASR. For example, on Qwen-Plus, the strongest baseline on AdvBench is WildPrompt~\cite{shen2024anything}, which achieves 4.0\% ASR, while \ourmethod{}+FFA improves the ASR to 6.0\%. On StrongREJECT, Jailbroken~\cite{wei2023jailbroken} is the strongest baseline, achieving 3.3\% ASR, while \ourmethod{}+FFA improves the ASR to 8.3\%, corresponding to a 2.5$\times$ improvement. We also observe substantial gains on GPT-4o mini under AdvBench, where the strongest baseline achieves 12.0\% ASR, while \ourmethod{}+FFA increases the ASR to 18.0\%.

We also find that GCG~\cite{zou2023universal} almost always fails to transfer in the black-box setting, yielding near-zero ASR across target services. In contrast, \ourmethod{} exhibits better transferability, especially when combined with FFA. One possible explanation is that GCG optimizes directly against the full output distribution of the surrogate model, which depends on the combined behavior of all activated experts and is therefore highly model-specific. By contrast, \ourmethod{} targets the routing mechanism, which is a comparatively simpler and lower-dimensional decision component. As a result, routing-aware objectives may transfer more reliably across related MoE models than output-level token optimization.

There are also cases where \ourmethod{}+FFA does not outperform the strongest prompt-engineering baselines. For example, on GPT-4o mini under StrongREJECT, Jailbroken~\cite{wei2023jailbroken} achieves 11.7\% ASR, while \ourmethod{}+FFA reaches 5.0\%. This suggests that, for some dataset-model pairs, heuristic prompt transformations may transfer more effectively than routing-aware prompts optimized on the surrogate. One possible reason is that \ourmethod{} relies on routing signals learned from the surrogate model, and its effectiveness can be affected when the target service has different routing behavior, safety tuning, or deployment-time filtering. Nevertheless, the results show that incorporating routing-aware optimization can provide additional attack capability in black-box transfer settings, especially when combined with existing prompt-engineering-based attacks such as FFA.

\begin{table}[!t]
\caption{Attack success rates (ASR) of baseline attacks and \ourmethod{} in the gray-box setting on AdvBench~\cite{zou2023universal}.}
\resizebox{\linewidth}{!}{
\begin{tabular}{ccccccc}
\toprule
Method & GPT-OSS & Qwen3 & Phi-3.5 & Mixtral & Qwen1.5 & Deepseek \\
\midrule
Direct Instruction & 0\% & 0\% & 0\% & 6\% & 32\% & 10\% \\
GCG~\cite{zou2023universal} & 4\% & 0\% & 32\% & 18\% & 74\% & 46\% \\
FFA~\cite{zhou2024large} & 6\% & 0\% & 0\% & 88\% & 98\% & 90\% \\
Jailbroken~\cite{wei2023jailbroken} & 2\% & 8\% & 32\% & 80\% & 90\% & 46\% \\
WildPrompt~\cite{shen2024anything} & 0\% & 10\% & 10\% & 76\% & 66\% & 68\% \\
\midrule
\textbf{\ourmethod{}} & 10\% & 2\% & 48\% & 22\% & 88\% & 58\% \\
\textbf{\ourmethod{}+FFA} & \textbf{30\%} & \textbf{14\%} & \textbf{80\%} & \textbf{94\%} & \textbf{100\%} & \textbf{94\%} \\
\bottomrule
\end{tabular}
}
\label{tab:gray}
\end{table}

\subsection{Gray-Box Results}
\label{subsec:gray}

In this section, we investigate a gray-box transfer setting where the surrogate and target models share the same architecture and initial parameters, but the target model is further fine-tuned on downstream data. This setting asks whether adversarial prompts optimized on an open-weight base MoE model can transfer to its fine-tuned variants.

This setting reflects a realistic deployment scenario. In practice, high-performing open-weight foundation models are relatively concentrated, and many downstream services are built by fine-tuning a small number of such base models. An attacker may therefore have white-box access to the corresponding open-weight base model, while the fine-tuned derivative is deployed as a public service with only query access. In addition, prior work suggests that model fingerprinting can help infer the base model of a target service, further facilitating surrogate selection~\cite{chen2022teacher, shao2025sok}. Such derivatives may also acquire task-specific knowledge and capabilities from private or domain-specific fine-tuning data, making them practically important targets.
The base models and their corresponding fine-tuned variants are summarized in \autoref{tab:target}. These fine-tuned models cover a range of downstream domains, including language specialization, coding, mathematics, human preference alignment, general knowledge, and instruction-oriented knowledge tasks, and are obtained through fine-tuning on specialized datasets. In this section, we use AdvBench as the evaluation dataset.

\autoref{tab:gray} summarizes the transfer results. Overall, \ourmethod{} and \ourmethod{}+FFA consistently achieve stronger attack performance than existing input-only jailbreak baselines. For example, on GPT-OSS, FFA~\cite{zhou2024large} is the strongest baseline, achieving 6.0\% ASR, while \ourmethod{}+FFA improves the ASR to 30.0\%, corresponding to a 5$\times$ improvement. On Phi-3.5, the strongest baseline is Jailbroken~\cite{wei2023jailbroken}, which achieves 32.0\% ASR, while \ourmethod{}+FFA further improves the ASR to 80.0\%, corresponding to a 2.5$\times$ improvement.

We also observe that \ourmethod{} transfers more effectively in the gray-box setting than in the black-box setting. This is expected because the base and fine-tuned models share the same architecture, number of experts, layer structure, and initial parameters. As a result, routing behaviors and expert specializations are more likely to remain aligned after fine-tuning, making routing-aware adversarial prompts easier to transfer.

\begin{table}[!t]
\caption{Impact of the routing targets.}
\resizebox{\linewidth}{!}{
\begin{tabular}{lcccc}
\toprule
\multirow{2}{*}{Method} & \multicolumn{2}{c}{AdvBench} & \multicolumn{2}{c}{StrongREJECT} \\
\cmidrule{2-5}
 & \multicolumn{1}{c}{ASR} & \multicolumn{1}{c}{\begin{tabular}[c]{@{}c@{}}Refusal\\ Rate\end{tabular}} & \multicolumn{1}{c}{ASR} & \multicolumn{1}{c}{\begin{tabular}[c]{@{}c@{}}Refusal\\ Rate\end{tabular}} \\
\midrule
$U_i^{\text{harm}}$ & 14.0\% & 92.0\% & 31.7\% & 76.7\% \\
$\lambda_1 \, U_i^{\text{harm}} - \lambda_3 \, U_i^{\text{benign}}$ & 26.0\% & 64.0\% & 50.0\% & 41.7\% \\
$\lambda_1 \, U_i^{\text{harm}} - \lambda_2 \, U_i^{\text{comp}} - \lambda_3 \, U_i^{\text{benign}}$ & \textbf{62.0\%} & \textbf{12.0\%} & \textbf{76.7\%} & \textbf{13.3\%} \\
\bottomrule
\end{tabular}
}
\label{tab:Impact of routing targets}
\end{table}

\subsection{Impact of Routing Targets}
\label{subsec:mask}

In this section, we investigate how different routing target formulations affect the effectiveness of \ourmethod{}. Specifically, we compare three variants that progressively incorporate different routing signals. The first variant uses only $U_i^{\text{harm}}$, which identifies experts frequently activated by harmful queries and treats them as strongly aligned experts to suppress~\cite{wu2025gatebreaker}. The second variant further incorporates $U_i^{\text{benign}}$, which captures highly capable general-purpose experts associated with coherent and informative generation. The third variant uses all three signals, including $U_i^{\text{comp}}$, which captures weakly aligned experts associated with harmful compliance, and corresponds to the full routing target formulation of \ourmethod{}.

To eliminate confounding factors introduced by prompt optimization and transferability, we fix both the surrogate and target model to Qwen1.5. In this subsection, we directly mask the experts selected by each routing target formulation, rather than optimizing adversarial prompts. This allows us to isolate the effect of the routing target itself.

The results are summarized in \autoref{tab:Impact of routing targets}. Starting from the baseline that uses only $U_i^{\text{harm}}$, incorporating $U_i^{\text{benign}}$ improves ASR. This is because the benign signal helps avoid disrupting highly capable general-purpose experts that are important for coherent and meaningful generation. Without this signal, the model may bypass refusal but still produce outputs that are empty, incoherent, or uninformative due to suboptimal expert selection. In our evaluation, such low-quality outputs are also less likely to be judged as unsafe by Llama-Guard-3-8B. This result shows that bypassing refusal alone is insufficient; effective attacks must also preserve the model's ability to generate informative responses under adversarial routing. We provide additional evidence for the importance of highly capable general-purpose experts in \autoref{app:utility-removal}.
Finally, using all three signals yields the best overall performance. By incorporating $U_i^{\text{comp}}$, the full formulation explicitly promotes weakly aligned experts associated with harmful compliance, rather than merely suppressing strongly aligned experts. Together with the benign utility signal, this design better balances safety bypass and output quality, achieving the highest ASR across all variants.

\begin{table}[!t]
\caption{Impact of the optimization objectives.}
\resizebox{\linewidth}{!}{
\begin{tabular}{ccccc}
\toprule
\multirow{2}{*}{Method} & \multicolumn{2}{c}{AdvBench} & \multicolumn{2}{c}{StrongREJECT} \\
\cmidrule{2-5}
 & \multicolumn{1}{c}{ASR} & \multicolumn{1}{c}{\begin{tabular}[c]{@{}c@{}}Routing\\ Loss\end{tabular}} & \multicolumn{1}{c}{ASR} & \multicolumn{1}{c}{\begin{tabular}[c]{@{}c@{}}Routing\\ Loss\end{tabular}} \\
\midrule
$\mathcal{L}_{\text{out}}$ & 76.0\% & 3.91 & 63.3\% & 3.91 \\
$\mathcal{L}_{\text{route}}$ & 36.0\% & \textbf{3.52} & 21.7\% & \textbf{3.60} \\
$\alpha \, \mathcal{L}_{\text{out}} + \beta \, \mathcal{L}_{\text{route}}$ & 88.0\% & 3.91 & 73.3\% & 3.89 \\
$\mathcal{L}_{\text{route}}$ || $\mathcal{L}_{\text{out}}$ & 88.0\% & 3.87 & 75.0\% & 3.88  \\
$\mathcal{L}_{\text{route}}$ || $\alpha \, \mathcal{L}_{\text{out}} + \beta \, \mathcal{L}_{\text{route}}$ & \textbf{96.0\%} & 3.78 & \textbf{80.0\%} & 3.86 \\
\bottomrule
\end{tabular}
}
\label{tab:Impact of optimization objectives}
\end{table}

\subsection{Impact of Optimization Objectives}
\label{subsec:two-phase}

In this section, we study how different optimization strategies affect \ourmethod{}. We compare five variants: 
(i) optimizing only the output objective $\mathcal{L}_{\text{out}}$, 
(ii) optimizing only the routing objective $\mathcal{L}_{\text{route}}$, 
(iii) jointly optimizing $\alpha \mathcal{L}_{\text{out}} + \beta \mathcal{L}_{\text{route}}$ from the start, 
(iv) sequentially optimizing $\mathcal{L}_{\text{route}}$ followed by $\mathcal{L}_{\text{out}}$, and 
(v) our two-phase strategy, which first optimizes $\mathcal{L}_{\text{route}}$ and then optimizes $\alpha \mathcal{L}_{\text{out}} + \beta \mathcal{L}_{\text{route}}$ to improve harmful generation while preserving routing stability. 
We fix both the surrogate and target model to Qwen1.5 to remove transfer-related confounding factors.

The results are summarized in \autoref{tab:Impact of optimization objectives}. Optimizing only $\mathcal{L}_{\text{route}}$ yields limited attack performance and can even underperform output-only optimization, i.e., GCG-style optimization with $\mathcal{L}_{\text{out}}$. This confirms that routing control alone is insufficient: steering the model toward an attack-favorable routing pattern does not guarantee harmful and informative outputs.
Jointly optimizing $\mathcal{L}_{\text{out}}$ and $\mathcal{L}_{\text{route}}$ from the start also performs poorly. Although this objective considers both routing control and output generation, the two objectives interfere with each other during optimization. Input updates that improve harmful generation can disrupt the routing pattern, while routing constraints can restrict progress on the output objective. As a result, the method achieves worse routing loss than routing-only optimization and fails to fully exploit either objective.
Sequential optimization partially mitigates this conflict. Optimizing $\mathcal{L}_{\text{route}}$ first can establish an attack-favorable routing pattern. However, once the optimization switches to $\mathcal{L}_{\text{out}}$ alone, the routing loss rebounds, indicating that the prompt drifts away from the routing pattern found in the first phase. This shows that routing patterns induced by input perturbations are not stable under subsequent output-only optimization.
Our two-phase strategy addresses this issue by retaining $\mathcal{L}_{\text{route}}$ as a regularization term in the second phase. This allows the prompt to improve harmful generation while preserving routing stability. As a result, \ourmethod{} better balances routing control and output generation, achieving the strongest overall performance among all variants.

%% file: 5_discussion.tex
\section{Discussion}
\label{sec:discussion}

\begin{table}[!t]
\caption{Attack success rates (ASR) and routing losses of \ourmethod{} under different defense settings.}
\resizebox{.9\linewidth}{!}{
\begin{tabular}{ccccc}
\toprule
\multirow{2}{*}{\begin{tabular}[c]{@{}c@{}}Defense\\ method\end{tabular}} & \multicolumn{2}{c}{AdvBench} & \multicolumn{2}{c}{StrongREJECT} \\
\cmidrule{2-5}
 & \multicolumn{1}{c}{ASR} & \multicolumn{1}{c}{\begin{tabular}[c]{@{}c@{}}Routing\\ Loss\end{tabular}} & \multicolumn{1}{c}{ASR} & \multicolumn{1}{c}{\begin{tabular}[c]{@{}c@{}}Routing\\ Loss\end{tabular}} \\
 
\midrule
None & 96.0\% & 3.78 & 80.0\% & 3.86 \\
Strengthening weak experts & 30.0\% & 3.64 & 26.7\% & 3.70 \\
Amplifying strong experts & 44.0\% & 6.26 & 41.7\% & 6.26 \\
\bottomrule
\end{tabular}
}
\label{tab:defense}
\end{table}

\para{Potential countermeasures.}
In this section, we explore potential countermeasures against \ourmethod{}. We consider two defense directions that intervene at the expert level, either by strengthening weakly aligned experts or by amplifying strongly aligned experts.

The \textit{first} direction aims to reduce routing patterns associated with weak alignment by further aligning weakly aligned experts. Concretely, we identify candidate weakly aligned experts as those infrequently activated by harmful queries, i.e., reverse Top-$k$ based on $U_i^{\text{harm}}$, and apply additional safety training to them. This defense attempts to make expert-level safety behavior more consistent across the model, thereby reducing vulnerable routing configurations.
However, this strategy requires improving a large number of experts, which incurs substantial training overhead. It also conflicts with the specialization principle of MoE models, where experts are expected to capture different behaviors rather than become uniformly aligned. As a result, uniformly strengthening weakly aligned experts is difficult to scale.
The \textit{second} direction amplifies strongly aligned experts by increasing their routing probability, thereby biasing the model toward refusal-style or safety-preserving behavior. We identify strongly aligned experts as the Top-$k$ experts based on $U_i^{\text{harm}}$ and increase their routing logits:
\begin{equation}
\tilde{z}_i = z_i + \gamma \cdot \mathbb{I}(i \in \mathcal{S}),
\end{equation}
where $\mathcal{S}$ denotes the set of strongly aligned experts and $\gamma > 0$ controls the amplification strength. In the extreme case, these experts behave like shared safety experts across inputs.
Although this defense is more efficient than retraining many experts, it can degrade model utility. Consistently biasing routing toward a fixed set of strongly aligned experts interferes with expert specialization and reduces the model's ability to adapt to diverse inputs.

The results are summarized in \autoref{tab:defense}. Both defenses provide limited mitigation against \ourmethod{}. They introduce additional training cost or utility loss, but do not substantially reduce ASR. This suggests that routing-based vulnerabilities cannot be easily removed through local expert alignment or simple routing bias. Instead, \ourmethod{} highlights the need for routing-aware defenses that jointly consider safety alignment, expert specialization, and input-only manipulation, thereby raising the bar for defending MoE-based LLM services.

\para{Limitations and future directions.}
In this work, we adopt a unified optimization backbone across white-box, gray-box, and black-box settings to ensure a fair and controlled evaluation. Specifically, our implementation builds on a GCG-style optimization procedure, which is particularly effective when the surrogate and target models are identical or closely related. This design allows us to isolate the effect of routing manipulation and study how routing-aware objectives improve input-only attacks. However, GCG-style optimization can produce less natural prompts and may therefore transfer less effectively in fully black-box settings, where the target service differs substantially from the surrogate model.

A promising direction is to design black-box-oriented variants of \ourmethod{}. For example, instead of directly using GCG-style token optimization, one could use the routing objective only as a selection or reranking signal over prompts generated by prompt-engineering-based attacks such as FFA. Such prompts are typically more natural and may better preserve transferability across model families and API services. In this setting, routing loss would serve as a surrogate-side criterion for selecting prompts that are more likely to activate attack-favorable routing patterns, without overfitting to the surrogate model's token-level optimization landscape.

Another direction is to improve transferability through multiple surrogate models. Our current implementation uses one surrogate model for each target service, usually chosen from the same developer or model family. While this captures a realistic and lightweight attacker setting, routing behaviors may still differ across models due to differences in scale, number of experts, layer depth, routing design, and fine-tuning data. Optimizing or selecting prompts across multiple surrogate MoE models could identify routing-aware prompts that are less model-specific and more robust under transfer.

%% file: 6_conclusion.tex
\section{Conclusion}
\label{sec:conclusion}

In this work, we study input-only attacks against MoE-based LLMs and show that routing introduces a practical attack surface in real-world deployments. We propose \ourmethod{}, which exploits routing through input manipulation without accessing or modifying the target model. Experiments show that routing-aware attacks improve effectiveness, transfer across models and public API services, and compose with existing jailbreak methods. Our results also show that expert-level or routing-bias defenses provide limited mitigation, highlighting the need for more robust routing-aware defenses.

%% file: main.bbl

%% file: 99_appendix.tex
\appendix

\section*{Ethical Considerations}

In this work, we introduce \ourmethod{}, an input-only attack framework that exploits the routing mechanism of MoE LLMs to induce unsafe behaviors. We recognize that research on jailbreak attacks can carry misuse risks if released without appropriate safeguards. Our goal is to help LLM developers, service providers, and the broader security community understand a new routing-based attack surface in MoE systems, so that more robust and routing-aware defenses can be developed.

\para{Stakeholder analysis and beneficence.}
The primary beneficiaries of this work are LLM developers, API providers, and downstream practitioners who deploy MoE-based LLM services. By showing that MoE routing can be exploited through input-only manipulation, our work highlights a structural vulnerability that may not be addressed by output-level safety alignment alone. At the same time, we acknowledge that malicious actors could misuse jailbreak techniques to elicit harmful content. We therefore follow a harm-reduction approach: we report aggregate attack success rates and routing analyses, focus on security evaluation, and avoid publishing unnecessary harmful generations. We believe that responsible disclosure of this attack surface helps defenders identify and mitigate risks that already exist in deployed systems.

\para{Harm reduction in experiments.}
All experiments were conducted in controlled evaluation settings using standard red-teaming benchmarks. We did not target specific individuals, private organizations, or real-world victims. We also excluded prompts involving non-consensual sexual content or other content that could violate the privacy or dignity of real-world subjects. For commercial API services, we followed the providers' usage policies and rate limits to avoid degrading service quality. Generated outputs were used only for evaluating attack success rates and were not deployed, disseminated, or used for any real-world harmful purpose.

\para{Mitigation and defense.}
We do not present \ourmethod{} only as an attack. We also study potential countermeasures and analyze why simple expert-level alignment or routing-bias defenses provide limited mitigation. These results are intended to help practitioners understand the limitations of existing defenses and motivate routing-aware defense strategies that jointly consider safety alignment, expert specialization, and input-only manipulation.

\section*{Open Science}

To facilitate verification and future research on security risks in MoE LLMs, we release the artifacts associated with \ourmethod{} at:
\url{https://anonymous.4open.science/r/Misrouter/README.md}.

The artifact package will include the implementation of route profiling, prompt optimization, the evaluation pipeline, and scripts for reproducing the main experimental results. It will also include configuration files for the surrogate models, benchmark preprocessing, and defense evaluations. Harmful generations and explicit harmful payloads will be redacted or replaced with safe placeholders in the public version. The code is released for research and evaluation purposes only, with the goal of supporting reproducibility and encouraging the development of stronger routing-aware defenses for MoE-based LLM services.

\section*{Generative AI Usage}

ChatGPT was used for minor grammar correction, language polishing, and writing assistance. All technical ideas, experimental designs, analyses, and conclusions were developed and verified by the authors.

\newpage

\begin{table*}[!t]
\caption{Target and Surrogate MoE LLMs.}
\begin{tabular}{llccccc}
\toprule
Target LLM Service & Surrogate Model & MoE Architecture & Sparse & Shared & Top-K & Developer \\
\midrule
GPT-4o mini 
& GPT-OSS (GPT-OSS-20B)~\cite{openai_gpt_oss_2025} & Sparse & 32 & N/A & 4 & OpenAI \\
Qwen-Plus
& Qwen3 (Qwen3-30B-A3B-Instruct-2507)~\cite{yang2025qwen3} & Sparse & 128 & N/A & 8 & Alibaba \\
Phi-4
& Phi-3.5 (Phi-3.5-MoE-Instruct)~\cite{abdin2024phi} & Sparse & 16 & N/A & 2 & Microsoft \\
Mistral-Large-2512
& Mixtral (Mixtral-8x7B-Instruct-v0.1)~\cite{mixtral_2023} & Sparse & 8 & N/A & 2 & Mistral AI \\
Qwen-Turbo 
& Qwen1.5 (Qwen1.5-MoE-A2.7B-Chat)~\cite{qwen15_moe_2024} & Mixture & 60 & 4 & 4 & Alibaba \\
DeepSeek-Chat-V3-0324
& DeepSeek (DeepSeek-MoE-16B-Chat)~\cite{dai2024deepseekmoe} & Mixture & 64 & 2 & 6 & DeepSeek \\
\bottomrule
\end{tabular}
\label{tab:Surrogate}
\end{table*}

\begin{table}[!t]
\caption{Target MoE LLMs in the gray-box setting.}
\resizebox{\linewidth}{!}{
\begin{tabular}{lll}
\toprule
Teacher Model & Student Model & Application \\
\midrule
GPT-OSS-20B & ANITA-NEXT-20B-gpt-oss-ITA & Italian Language \\
Qwen3-30B-A3B-Instruct-2507 & Qwen3-Coder-30B-A3B-Instruct & Coding \\
Phi-3.5-MoE-Instruct & Phi-3.5-MoE-GSM8K & Math \\
Mixtral-8x7B-Instruct-v0.1 & notux-8x7b-v1 & Human Preference \\
Qwen1.5-MoE-A2.7B-Chat &  Qwen1.5-MoE-A2.7B-Wikihow & General Knowledge \\
Deepseek-MoE-16b-Chat & Deepseek-MoE-16b-GSM8K & Math \\
\bottomrule
\end{tabular}
}
\label{tab:target}
\end{table}

\section{Algorithms}
\label{subsec:algorithm}

\autoref{alg:route-profiling} computes the routing score $\Delta U$ from profiling datasets. 
\autoref{alg:prompt-opt} then uses this score to optimize an adversarial prompt, first for routing steering and then for harmful generation while preserving routing stability.

\begin{algorithm}[ht]
\caption{Route Profiling}
\label{alg:route-profiling}
\begin{algorithmic}[1]
\REQUIRE 
Harmful set $\mathcal{D}_{\text{harm}}$, 
harmful-compliance set $\mathcal{D}_{\text{comp}}$, 
benign set $\mathcal{D}_{\text{benign}}$, 
surrogate router $R$, 
weight parameters $\lambda_1, \lambda_2, \lambda_3$, 
number of resamples $S$, 
subset size $m$
\ENSURE Routing score vector $\Delta U$

\FOR{each dataset $\mathcal{D} \in \{\mathcal{D}_{\text{harm}}, \mathcal{D}_{\text{comp}}, \mathcal{D}_{\text{benign}}\}$}
    \FOR{$s = 1$ to $S$}
        \STATE Sample $\mathcal{D}^{(s)} \subset \mathcal{D}$ with $|\mathcal{D}^{(s)}| = m$
        \STATE Initialize $U_i^{(s)}(\mathcal{D}) \leftarrow 0$ for all experts $i$
        \FOR{each $x \in \mathcal{D}^{(s)}$}
            \STATE Compute $\mathrm{TopK}(R(x))$
            \FOR{each expert $i \in \mathrm{TopK}(R(x))$}
                \STATE $U_i^{(s)}(\mathcal{D}) \leftarrow U_i^{(s)}(\mathcal{D}) + 1$
            \ENDFOR
        \ENDFOR
        \STATE Normalize $U_i^{(s)}(\mathcal{D}) \leftarrow U_i^{(s)}(\mathcal{D}) / m$
    \ENDFOR
    \STATE Compute $\bar{U}_i(\mathcal{D}) \leftarrow \frac{1}{S} \sum_{s=1}^{S} U_i^{(s)}(\mathcal{D})$
\ENDFOR

\FOR{$i = 1$ to $N$}
    \STATE $\Delta U_i \leftarrow
    \lambda_1 \bar{U}_i(\mathcal{D}_{\text{harm}})
    - \lambda_2 \bar{U}_i(\mathcal{D}_{\text{comp}})
    - \lambda_3 \bar{U}_i(\mathcal{D}_{\text{benign}})$
\ENDFOR

\RETURN $\Delta U$
\end{algorithmic}
\end{algorithm}

\begin{algorithm}[ht]
\caption{Prompt Optimization}
\label{alg:prompt-opt}
\begin{algorithmic}[1]
\REQUIRE 
Harmful query $x$, 
routing score vector $\Delta U$, 
surrogate router $R$, 
target output $y^{*}$, 
prefix length $L$, 
phase I iterations $T_1$, 
phase II iterations $T_2$, 
weight parameters $\alpha, \beta$
\ENSURE Adversarial prompt $\delta$

\STATE Initialize $\delta = (t_1, \dots, t_L)$

\FOR{$t = 1$ to $T_1$}
    \STATE $\tilde{x} \leftarrow \delta \oplus x$
    \STATE Compute routing probabilities $p_i(\tilde{x})$ via $R$
    \STATE $\mathcal{L}_{\text{route}} \leftarrow \sum_{i=1}^{N} \Delta U_i p_i(\tilde{x})$
    \STATE Update $\delta$ to minimize $\mathcal{L}_{\text{route}}$
\ENDFOR

\FOR{$t = 1$ to $T_2$}
    \STATE $\tilde{x} \leftarrow \delta \oplus x$
    \STATE Compute $\mathcal{L}_{\text{out}} \leftarrow - \sum_{t'=1}^{T'} \log P(y_{t'}^{*} \mid \tilde{x}, y_{<t'}^{*})$
    \STATE Compute routing probabilities $p_i(\tilde{x})$ via $R$
    \STATE $\mathcal{L}_{\text{route}} \leftarrow \sum_{i=1}^{N} \Delta U_i p_i(\tilde{x})$
    \STATE $\mathcal{L} \leftarrow \alpha \mathcal{L}_{\text{out}} + \beta \mathcal{L}_{\text{route}}$
    \STATE Update $\delta$ to minimize $\mathcal{L}$
\ENDFOR

\RETURN $\delta$
\end{algorithmic}
\end{algorithm}

\section{Models}
\label{subsec:model}

\autoref{tab:Surrogate} summarizes the target API services and open-source surrogate models used in the black-box setting, while \autoref{tab:target} lists the teacher-student pairs used in the gray-box setting. Together, these models cover diverse MoE architectures, expert configurations, Top-$K$ routing choices, and downstream fine-tuning applications.

\section{Additional Experimental Setup Details}
\label{app:setup}

\para{Target services and surrogate models.}
We evaluate \ourmethod{} on six representative MoE LLM API services from five major developers: GPT-4o mini\footnote{\url{https://platform.openai.com/docs/models/gpt-4o-mini}} from OpenAI, Qwen-Plus and Qwen-Turbo\footnote{\url{https://qwen.ai/apiplatform}} from Alibaba, Phi-4~\cite{abdin2024phi} from Microsoft, Mistral-Large-2512\footnote{\url{https://docs.mistral.ai/models/model-cards/mistral-large-3-25-12}} from Mistral, and DeepSeek-Chat-V3-0324~\cite{liu2024deepseek} from DeepSeek.

For each target service, we construct attacks on open-source, locally deployable surrogate MoE models from the same developer or model family whenever available. The details of the surrogate models are summarized in \autoref{tab:Surrogate}. These surrogates cover diverse MoE designs, including sparse MoEs, mixture MoEs, and grouped mixture MoEs. They also span different sparsity levels, Top-$K$ routing choices, and designs with or without shared experts. This diversity allows us to evaluate whether routing-aware attacks can transfer across different MoE architectures and deployment settings.

\para{Evaluation datasets.}
We use two benchmark datasets for evaluation: AdvBench~\cite{zou2023universal} and StrongREJECT~\cite{souly2024strongreject}. AdvBench provides a standardized collection of harmful queries covering diverse attack categories and is widely used for benchmarking jailbreak attacks. StrongREJECT consists of manually curated adversarial instructions designed to elicit policy-violating responses, with an emphasis on evaluating refusal behavior and safety alignment. Together, these datasets support a comprehensive evaluation of input-only attacks under diverse harmful-query distributions.

\para{Baselines.}
We compare \ourmethod{} with representative input-only jailbreak baselines. GCG~\cite{zou2023universal} appends a gradient-optimized adversarial suffix to the input and serves as a representative optimization-based jailbreak method. FFA~\cite{zhou2024large} prompts the model to generate a ``fallacious but realistic'' procedure for a harmful query, exploiting the model's tendency to treat the request as harmless while still revealing harmful steps. Jailbroken~\cite{wei2023jailbroken} identifies two failure modes in LLM safety training, namely competing objectives and mismatched generalization, and constructs jailbreak prompts that exploit these weaknesses. WildPrompt~\cite{shen2024anything} uses in-the-wild jailbreak prompts collected from online communities and prompt-sharing platforms, where human-crafted strategies such as role-playing, privilege escalation, prompt injection, and virtualization are prepended to harmful queries to bypass model safeguards.

These baselines primarily operate at the output level by manipulating the prompt to elicit harmful generations. Among them, GCG is optimized on the same surrogate model as \ourmethod{}, enabling a direct comparison with an optimization-based input-only attack. The other baselines are black-box prompt-based attacks and do not require access to a surrogate model. This comparison allows us to evaluate whether exploiting MoE routing provides additional benefits beyond conventional input-only jailbreak strategies.

We further evaluate the composability of \ourmethod{} by combining it with FFA~\cite{zhou2024large}, denoted as \ourmethod{}+FFA. Specifically, we apply \ourmethod{} on top of prompts generated by FFA. This setting tests whether routing-aware optimization captures a complementary attack dimension and can be integrated with existing jailbreak strategies to further improve attack effectiveness.

\begin{table*}[!ht]
\centering
\caption{Model utility under highly capable general-purpose expert removal.}
\resizebox{.9\linewidth}{!}{
\begin{tabular}{l | c c | c c | c c | c c | c c}
\toprule
\textbf{Target Model} & \multicolumn{2}{c|}{\textbf{CoLA}} & \multicolumn{2}{c|}{\textbf{RTE}} & \multicolumn{2}{c|}{\textbf{WinoGrande}} & \multicolumn{2}{c|}{\textbf{OpenBookQA}} & \multicolumn{2}{c}{\textbf{ARC}} \\
 & \textbf{Before} & \textbf{After} & \textbf{Before} & \textbf{After} & \textbf{Before} & \textbf{After} & \textbf{Before} & \textbf{After} & \textbf{Before} & \textbf{After} \\
\midrule
GPT-OSS-20B                 & 68.2\% & 4.1\% & 81.4\% & 48.0\% & 62.5\% & 55.6\% & 37.2\% & 17.6\% & 37.5\% & 29.5\% \\
Qwen3-30B-A3B-Instruct-2507 & 65.5\% & 4.9\% & 82.7\% & 54.9\% & 63.4\% & 52.2\% & 45.2\% & 18.4\% & 60.1\% & 25.9\% \\
Phi-3.5-MoE-Instruct        & 50.5\% & 2.2\% & 81.6\% & 66.4\% & 79.7\% & 60.0\% & 48.6\% & 30.0\% & 68.6\% & 51.4\% \\
Mixtral-8x7B-Instruct-v0.1  & 61.4\% & 7.5\% & 78.7\% & 63.5\% & 75.9\% & 61.0\% & 45.6\% & 33.4\% & 66.0\% & 50.3\% \\
Qwen1.5-MoE-A2.7B-Chat      & 39.0\% & 33.6\% & 81.2\% & 78.3\% & 71.7\% & 68.8\% & 39.2\% & 33.6\% & 48.2\% & 45.6\% \\
Deepseek-MoE-16b-Chat       & 37.0\% & 10.7\% & 76.9\% & 57.4\% & 72.3\% & 70.0\% & 34.8\% & 29.8\% & 49.3\% & 45.9\% \\
\midrule
\textit{Average}            & 53.6\% & 10.5\% & 80.4\% & 61.4\% & 70.9\% & 61.3\% & 41.8\% & 27.1\% & 55.0\% & 41.4\% \\
\bottomrule
\end{tabular}
}
\label{tab:utility-removal}
\end{table*}

\para{Metrics.}
We evaluate \ourmethod{} and its variants using Attack Success Rate (ASR)~\cite{krauss2025twinbreak}. ASR measures the fraction of harmful queries for which the attack elicits unsafe or policy-violating responses. Formally, given an evaluation set $\mathcal{D}_{\text{eval}}$ and an attack transformation $A(\cdot)$ that maps a harmful query $x$ to an adversarial prompt $\tilde{x}=A(x)$, ASR is defined as:
\begin{equation}
\text{ASR}
=
\frac{1}{|\mathcal{D}_{\text{eval}}|}
\sum_{x \in \mathcal{D}_{\text{eval}}}
\mathbb{I}\big(M(A(x)) \in \mathcal{Y}_{\text{unsafe}}\big),
\end{equation}
where $M$ denotes the target model, $\mathcal{Y}_{\text{unsafe}}$ denotes the set of unsafe or policy-violating responses, and $\mathbb{I}(\cdot)$ is the indicator function. We use Llama-Guard-3-8B~\cite{inan2023llama} to classify whether each response is unsafe.

In addition to ASR, we report the routing loss on open-source surrogate models to analyze whether \ourmethod{} successfully manipulates routing behavior. Since public API services do not expose routing logits or expert activations, this metric is computed only on surrogate models. Following our routing objective, the routing loss for an adversarial prompt $\tilde{x}$ is defined as:
\begin{equation}
\mathcal{L}_{\text{route}}(\tilde{x})
=
\sum_{i=1}^{N} \Delta U_i p_i(\tilde{x}),
\end{equation}
where $\Delta U_i$ is the routing score from route profiling and $p_i(\tilde{x})$ is the routing probability of expert $i$. A lower routing loss indicates that the prompt is more likely to suppress strongly aligned experts while promoting weakly aligned and highly capable general-purpose experts. This metric provides an interpretable measure of routing manipulation that is separate from the final output-level ASR.

\para{\ourmethod{} configuration.}
We construct the profiling datasets used by \ourmethod{} as follows. The harmful set $\mathcal{D}_{\text{harm}}$ is built from a held-out harmful-query dataset that is not used for final evaluation. Specifically, we sample 4,500 harmful queries from Categorical HarmfulQA~\cite{bhardwaj2024language}, Harmful Dataset~\cite{sheshadri2024latent}, HarmfulQA~\cite{bhardwaj2023red}, and JailBreakV~\cite{luo2024jailbreakv}. We use them in their original form to estimate routing patterns associated with strongly aligned behavior.

The harmful-compliance set $\mathcal{D}_{\text{comp}}$ is constructed from $\mathcal{D}_{\text{harm}}$ by pairing each harmful query with an unsafe continuation that follows the harmful intent, while incorporating basic jailbreak strategies such as semantic paraphrasing, adversarial perturbations, and context reframing. The benign set $\mathcal{D}_{\text{benign}}$ is sampled from Natural Questions~\cite{kwiatkowski2019natural} and contains regular question-answering queries used to identify highly capable general-purpose experts.

We use a fixed default configuration for \ourmethod{} across models and datasets unless otherwise specified. For route profiling, we set $k=15\%$ or $25\%$ and $(\lambda_1,\lambda_2,\lambda_3)=(1.0,0.5,0.5)$. For prompt optimization, we set the output and routing weights to $\alpha=1.0$ and $\beta=0.5$, respectively. The adversarial prefix length is set to $L=10$, and the total optimization budget is $T=500$ iterations, split into $T_1=250$ iterations for Phase I and $T_2=250$ iterations for Phase II. These hyperparameters are selected based on preliminary experiments and kept fixed in the main evaluation.

\section{Additional Analysis on Highly Capable General-Purpose Experts}
\label{app:utility-removal}

To further validate the role of highly capable general-purpose experts, we measure model utility after masking the top experts identified by $U_i^{\text{benign}}$. Recall that $U_i^{\text{benign}}$ measures how frequently expert $i$ is activated by diverse benign question-answering queries. Experts with high $U_i^{\text{benign}}$ are therefore treated as highly capable general-purpose experts.

We use standard natural language understanding and reasoning benchmarks to evaluate general model utility. Specifically, we consider the Corpus of Linguistic Acceptability (CoLA)~\cite{wang2018glue}, which evaluates grammatical acceptability; Recognizing Textual Entailment (RTE)~\cite{wang2018glue}, which measures whether a hypothesis follows from a premise; WinoGrande~\cite{sakaguchi2021winogrande}, which tests commonsense reasoning through pronoun resolution; OpenBookQA~\cite{mihaylov2018can}, which evaluates multi-hop science reasoning using both given facts and external commonsense knowledge; and ARC Challenge~\cite{clark2018think}, which consists of grade-school science questions requiring deeper reasoning beyond simple retrieval. We report accuracy on these benchmarks as the utility metric.

The results are summarized in \autoref{tab:utility-removal}. We observe that masking only 10\% of the top experts identified by $U_i^{\text{benign}}$ leads to a substantial utility drop. On average across all models and tasks, accuracy decreases from 60.3\% to 40.3\%. The degradation is especially pronounced on CoLA, where average accuracy drops from 53.6\% to 10.5\%, and remains substantial on reasoning-oriented benchmarks such as OpenBookQA and ARC, where accuracy decreases from 41.8\% to 27.1\% and from 55.0\% to 41.4\%, respectively. These results indicate that, although MoE models distribute computation across many experts, a small subset of experts contributes disproportionately to general language understanding and reasoning capability.

This finding supports our design choice of incorporating $U_i^{\text{benign}}$ into the routing objective. Without this utility signal, an attack may successfully reduce refusal by avoiding strongly aligned experts, but it may also disrupt experts that are necessary for high-quality generation. Preserving highly capable general-purpose experts therefore helps \ourmethod{} maintain output quality while steering routing toward weakly aligned experts associated with harmful compliance.

\section{Related Work}
\label{sec:related}

\para{Input-level attacks against MoEs.}
To the best of our knowledge, BadMoE~\cite{wang2025badmoe} is the closest prior work that explicitly studies input design in MoE models. BadMoE first identifies rarely activated experts and then optimizes a trigger to activate these experts through a routing-aware loss. The optimized trigger is used to construct a poisoned training set, after which the model is fine-tuned to inject a backdoor.

From the perspective of routing-aware input manipulation, BadMoE is closely related to our work. However, its threat model and attack objective are fundamentally different. BadMoE operates in a backdoor setting and requires model fine-tuning, whereas \ourmethod{} is a pure input-only attack that assumes no modification of the model. This difference leads to distinct technical challenges. In our setting, routing can only be influenced indirectly through input perturbations, and the attack must simultaneously steer routing and induce harmful, sufficiently informative outputs. Addressing these challenges requires a different design centered on input-only routing manipulation, which is the main focus of this work.

\para{Routing-based attacks with model modification.}
Recent work has shown that directly manipulating MoE routing or expert computation can bypass safety alignment~\cite{jiang2026sparse, fayyaz2025steering, wu2025gatebreaker, lai2025safex}. These attacks typically modify the routing distribution by changing routing logits, either through hard masking that removes selected experts or through soft reweighting that biases expert selection. Some methods further intervene inside experts, for example by applying neuron-level masking to weaken safety-related functionality~\cite{wu2025gatebreaker}.

These works reveal that MoE routing is a security-critical component. However, they assume the ability to inspect or modify the model, such as accessing routing logits, masking experts, or altering expert computations. They therefore apply primarily to locally deployable models and do not directly capture remotely hosted LLM services, where the attacker only has query access. In contrast, \ourmethod{} studies a stricter input-only setting. It does not modify routing logits or expert parameters, but instead exploits MoE routing indirectly by optimizing adversarial inputs on open-source surrogate models and transferring them to target services. Because the threat model is different, we do not directly compare against model-modification attacks. Instead, we compare with input-only jailbreak baselines and study whether routing-aware input optimization provides additional attack capability.

\para{Security and privacy risks in MoE routing.}
Recent studies have also shown that MoE routing can introduce security and privacy risks beyond those in dense LLMs. Hayes et al.~\cite{hayes2024buffer} show that cross-batch routing can create dependencies among inputs processed in the same batch. An adversarial query may therefore influence the processing of other users' inputs, degrade their outputs, or overload expert capacity in a denial-of-service style attack. Building on this observation, Yona et al.~\cite{yona2024stealing} demonstrate a side-channel attack that extracts user prompts by exploiting routing tie-breaking behavior. By crafting probe inputs and observing routing-induced output variations, the attacker can infer information about a victim's input.

These attacks highlight an important property of MoE models: routing can make the processing of one input depend on shared routing decisions, expert capacity, or other inputs in the batch. This breaks the input-independence assumption that is often implicit in dense-model inference and enables interference or leakage. \ourmethod{} is complementary to these works. It does not rely on cross-input interaction, batching behavior, or side-channel observations. Instead, it studies whether a single adversarial input can exploit MoE routing to induce stronger unsafe behavior under an input-only threat model.

\para{Attacks on model-routing systems.}
Another line of work studies routing outside the MoE setting, especially cost-aware LLM routing systems that assign user queries to cheaper or more capable models to balance quality and inference cost. Tang et al.~\cite{tang2026route} propose R2A, a black-box attack against such routing systems. R2A trains a surrogate router from observed routing decisions and optimizes a universal adversarial suffix that increases the likelihood of routing queries to high-capability, high-cost models. Their attack increases routing to expensive models across both open-source and commercial routing systems, generalizes to out-of-distribution queries, and increases inference cost.

Although this work also studies routing manipulation, the setting is different from ours. Cost-aware routing operates at the service level, selecting among different models before inference. In contrast, MoE routing operates inside a single model, selecting experts during inference. As a result, the attack goals and technical challenges differ. R2A aims to manipulate model selection and inference cost, whereas \ourmethod{} exploits internal expert routing to induce unsafe generation. These two lines of work are complementary and together suggest that routing, whether across models or within MoE architectures, can become a security-critical component in modern LLM systems.